\documentclass[
aps,pre,
showkeys,
longbibliography,
reprint,                
twocolumn,              
nofootinbib,
floatfix
]{revtex4-1}

\usepackage{amsmath}
\usepackage{amsfonts}
\usepackage{graphicx}
\usepackage{tikz}
\usepackage{subcaption}

\usepackage{listings}
\definecolor{codegreen}{rgb}{0,0.6,0}
\definecolor{codegray}{rgb}{0.5,0.5,0.5}
\definecolor{codepurple}{rgb}{0.58,0,0.82}
\definecolor{backcolour}{rgb}{0.95,0.95,0.92}
\lstdefinestyle{mystyle}{
    backgroundcolor=\color{backcolour},   
    commentstyle=\color{codegreen},
    keywordstyle=\color{magenta},
    numberstyle=\tiny\color{codegray},
    stringstyle=\color{codepurple},
    basicstyle=\ttfamily\footnotesize,
    breakatwhitespace=false,         
    breaklines=true,                 
    captionpos=b,                    
    keepspaces=true,                 
    numbers=left,                    
    numbersep=5pt,                  
    showspaces=false,                
    showstringspaces=false,
    showtabs=false,                  
    tabsize=2
}
\usepackage{psfrag}
\usepackage{lipsum}

\usepackage[colorlinks=true,hyperfootnotes=true,breaklinks=true]{hyperref}
\usepackage[capitalise]{cleveref}

\begin{document}

\title{Random site percolation on honeycomb lattices with complex neighborhoods}

\author{Krzysztof Malarz}
\thanks{\includegraphics[width=10pt]{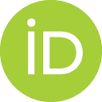}~\href{https://orcid.org/0000-0001-9980-0363}{0000-0001-9980-0363}}
\email{malarz@agh.edu.pl}
\affiliation{\mbox{AGH University, Faculty of Physics and Applied Computer Science},\\
al. Mickiewicza 30, 30-059 Krak\'ow, Poland}

\begin{abstract}
We present a rough estimation---up to four significant digits, based on the scaling hypothesis and the probability of belonging to the largest cluster vs. the occupation probability---of the critical occupation probabilities for the random site percolation problem on a honeycomb lattice with complex neighborhoods containing sites up to the fifth coordination zone.  
There are 31 such neighborhoods with their radius ranging from one to three and containing from three to 24 sites.
For two-dimensional regular lattices with compact extended-range neighborhoods, in the limit of the large number $z$ of sites in the neighborhoods, the site percolation thresholds $p_c$ follow the dependency $p_c\propto 1/z$, as recently shown by Xun, Hao and Ziff [\href{http://doi.org/10.1103/PhysRevE.105.024105}{Physical Review E {\bf 105}, 024105 (2022)}].
On the contrary, noncompact neighborhoods (with holes) destroy this dependence due to the degeneracy of the percolation threshold (several values of $p_c$ corresponding to the same number $z$ of sites in the neighborhoods).
An example of a single-value index $\zeta=\sum_i z_i r_i$---where $z_i$ and $r_i$ are the number of sites and radius of the $i$-th coordination zone, respectively ----characterizing the neighborhood and allowing avoiding the above-mentioned degeneracy is presented.
The percolation threshold obtained follows the inverse square root dependence $p_c\propto 1/\sqrt\zeta$. 
The functions {\tt boundaries()} (written in C) for basic neighborhoods (for the unique coordination zone) for the Newman and Ziff algorithm [\href{https://doi.org/10.1103/PhysRevE.64.016706}{Physical Review E {\bf 64}, 016706 (2001)}] are also presented.
\end{abstract}

\date{\today}

\maketitle

\section{Introduction}

The percolation thresholds \cite{Broadbent1957,Hammersley1957,bookDS,Wierman2014} (see Wikipedia page \cite{wiki.Percolation_threhold} for the most current and comprehensive data set and Refs.~\onlinecite{Li_2021,Saberi2015} for recent reviews) were initially estimated for the nearest neighbor interactions \cite{Dean_1963,Dean_Bird_1967,PhysRevE.60.275} but later also complex neighborhoods (called extended in case of compact one) were studied for 2D (square \cite{Dalton_1964,Domb1966,Gouker1983,Galam2005a,Galam2005b,Majewski2007,2010.02895}, 
triangular \cite{Dalton_1964,Domb1966,Iribarne1999,2006.15621,2102.10066,PhysRevE.105.024105}, 
honeycomb \cite{Dalton_1964,PhysRevE.105.024105}
and other Archimedean \cite{PhysRevE.105.024105}),
3D (simple cubic \cite{Kurzawski2012,Malarz2015,2010.02895,PhysRevE.105.024105})
and 4D (simple hypercubic \cite{1803.09504}) lattices.
As the exact values of $p_c$ are known only in several cases \cite{PhysRevE.105.044108} the most of effort in their calculations is computational.

Very recently, Xun, Hao, and Ziff \cite{PhysRevE.105.024105} numerically estimated the site and bond percolation thresholds for all eleven Archimedean lattices with extended compact neighborhoods containing sites up to the tenth coordination zone.
They found that for the site percolation problem, the critical site occupation probability $p_c$ follows asymptotically
\begin{equation}
\label{eq:pc-discs}
p_c= a/z
\end{equation}
with the total number $z$ of sites in the neighborhood and $a\approx 4.51235$.
This dependence should be reached exactly for the percolation of discs, that is, for compact neighborhoods with a large number $z$ of sites that make up the neighborhood.
The finite-$z$ effect may be taken into account by an additional including term $b$ in the denominator of \Cref{eq:pc-discs} \cite{PhysRevE.103.022127}
\begin{equation}
\label{eq:pc-finite-z}
p_c= c/(z+b)
\end{equation}
with $b=3$ for the two-dimensional lattices \cite{PhysRevE.105.024105}.
The third universal scaling studied in Ref.~\onlinecite{PhysRevE.105.024105} was
\begin{equation}
\label{eq:pc-exp}
p_c= 1-\exp(d/z)
\end{equation}
proposed by Koza et al. \cite{Koza_2014,Koza_2016}.

\begin{figure*}[htbp]
\begin{subfigure}[b]{0.19\textwidth}
\caption{\label{fig:1nn}}
\begin{tikzpicture}[scale=0.42]
\clip (2,1) rectangle (10,9);
\draw[line width=3mm,yellow,draw opacity=.5] 
(2,11*sin{60})--(10,6*sin{60})
(2,9*sin{60})--(10,4*sin{60}) (2,7*sin{60})--(10,2*sin{60});
  \foreach \i in {0,...,3} 
  \foreach \j in {0,...,5} {
  \foreach \a in {0,120,-120} \draw (3*\i,2*sin{60}*\j) -- +(\a:1);
  \foreach \a in {0,120,-120} \draw (3*\i+3*cos{60},2*sin{60}*\j+sin{60}) -- +(\a:1);}
\draw[ultra thick,red] (6, 6*sin{60}) circle (1);
\draw[ultra thick,green] (8.5, 5*sin{60}) circle (1);
\draw[very thick,red]   (6, 6*sin{60}) circle (7pt);
\draw[very thick,green] (8.5, 5*sin{60}) circle (7pt);
\filldraw[red] (7, 6*sin{60}) circle (7pt);
\filldraw[red] (5.5, 5*sin{60}) circle (7pt);
\filldraw[red] (5.5, 7*sin{60}) circle (7pt);
\filldraw[green] (7.5, 5*sin{60}) circle (7pt);
\filldraw[green] (9, 4*sin{60}) circle (7pt);
\filldraw[green] (9, 6*sin{60}) circle (7pt);
\end{tikzpicture}
\end{subfigure}
\hfill
\begin{subfigure}[b]{0.19\textwidth}
\caption{\label{fig:2nn}}
\begin{tikzpicture}[scale=0.42]
\clip (2,1) rectangle (10,9);
\draw[line width=3mm,yellow,draw opacity=.5] 
(2,11*sin{60})--(10,6*sin{60})
(2, 9*sin{60})--(10,4*sin{60})
(2, 7*sin{60})--(10,2*sin{60});
  \foreach \i in {0,...,3} 
  \foreach \j in {0,...,5} {
  \foreach \a in {0,120,-120} \draw (3*\i,2*sin{60}*\j) -- +(\a:1);
  \foreach \a in {0,120,-120} \draw (3*\i+3*cos{60},2*sin{60}*\j+sin{60}) -- +(\a:1);}
\draw[ultra thick] (6, 6*sin{60}) circle (2*sin{60});
\draw[very thick]     (6  , 6*sin{60}) circle (7pt);
\filldraw (6  , 8*sin{60}) circle (7pt);
\filldraw (6  , 4*sin{60}) circle (7pt);
\filldraw (4.5, 7*sin{60}) circle (7pt);
\filldraw (7.5, 7*sin{60}) circle (7pt);
\filldraw (4.5, 5*sin{60}) circle (7pt);
\filldraw (7.5, 5*sin{60}) circle (7pt);

\draw[very thick,dotted,blue] (6  , 8*sin{60})-- (4.5, 7*sin{60})-- (4.5, 5*sin{60})-- (6  , 4*sin{60})-- (7.5, 5*sin{60})-- (7.5, 7*sin{60})--(6  , 8*sin{60});

\draw[very thick,dotted,blue]
(6  , 6*sin{60})--(4.5, 7*sin{60})
(6  , 6*sin{60})--(4.5, 5*sin{60})
(6  , 6*sin{60})--(6  , 4*sin{60}) 
(6  , 6*sin{60})--(7.5, 5*sin{60})
(6  , 6*sin{60})--(7.5, 7*sin{60}) 
(6  , 6*sin{60})--(6  , 8*sin{60});

\end{tikzpicture}
\end{subfigure}
\hfill
\begin{subfigure}[b]{0.19\textwidth}
\caption{\label{fig:3nn}}
\begin{tikzpicture}[scale=0.42]
\clip (2,1) rectangle (10,9); 
\draw[line width=3mm,yellow,draw opacity=.5] 
(2,11*sin{60})--(10,6*sin{60})
(2, 9*sin{60})--(10,4*sin{60})
(2, 7*sin{60})--(10,2*sin{60});
  \foreach \i in {0,...,3} 
  \foreach \j in {0,...,5} {
  \foreach \a in {0,120,-120} \draw (3*\i,2*sin{60}*\j) -- +(\a:1);
  \foreach \a in {0,120,-120} \draw (3*\i+3*cos{60},2*sin{60}*\j+sin{60}) -- +(\a:1);}
\draw[ultra thick,red]   (6, 6*sin{60}) circle (2);
\draw[ultra thick,green] (7, 6*sin{60}) circle (2);
\draw[very thick,red]   (6, 6*sin{60}) circle (7pt);
\draw[very thick,green] (7, 6*sin{60}) circle (7pt);
\filldraw[red] (7, 8*sin{60}) circle (7pt);
\filldraw[red] (7, 4*sin{60}) circle (7pt);
\filldraw[red] (4, 6*sin{60}) circle (7pt);
\filldraw[green] (6, 8*sin{60}) circle (7pt);
\filldraw[green] (6, 4*sin{60}) circle (7pt);
\filldraw[green] (9, 6*sin{60}) circle (7pt);

\draw[very thick,dotted,red]
(6, 6*sin{60})--(7, 8*sin{60})--(9, 8*sin{60})--(10, 6*sin{60})
(6, 6*sin{60})--(7, 4*sin{60})--(9, 4*sin{60})--(10, 6*sin{60})
(6, 6*sin{60})--(4, 6*sin{60})
(7, 8*sin{60})--(6,10*sin{60})--(4,10*sin{60})--(3, 8*sin{60})--(4, 6*sin{60})
(7, 4*sin{60})--(6, 2*sin{60})--(4, 2*sin{60})--(3, 4*sin{60})--(4, 6*sin{60});
\end{tikzpicture}
\end{subfigure}
\hfill
\begin{subfigure}[b]{0.19\textwidth}
\caption{\label{fig:4nn}}
\begin{tikzpicture}[scale=0.42]
\clip (2,1) rectangle (10,9);
\draw[line width=3mm,yellow,draw opacity=.5] 
(2,13*sin{60})--(10,8*sin{60})
(2,11*sin{60})--(10,6*sin{60})
(2, 9*sin{60})--(10,4*sin{60})
(2, 7*sin{60})--(10,2*sin{60})
(2, 5*sin{60})--(10,0*sin{60});
  \foreach \i in {0,...,3} 
  \foreach \j in {0,...,5} {
  \foreach \a in {0,120,-120} \draw (3*\i,2*sin{60}*\j) -- +(\a:1);
  \foreach \a in {0,120,-120} \draw (3*\i+3*cos{60},2*sin{60}*\j+sin{60}) -- +(\a:1);}
\draw[ultra thick,red]   (6, 6*sin{60}) circle (3*sin{60});
\draw[ultra thick,green] (7, 6*sin{60}) circle (3*sin{60});
\draw[very thick,red]   (6, 6*sin{60}) circle (7pt);
\draw[very thick,green] (7, 6*sin{60}) circle (7pt);
\filldraw[red]   (8.5, 7*sin{60}) circle (7pt);
\filldraw[red]   (8.5, 5*sin{60}) circle (7pt);
\filldraw[red]   (5.5, 9*sin{60}) circle (7pt);
\filldraw[red]   (4.0, 8*sin{60}) circle (7pt);
\filldraw[red]   (5.5, 3*sin{60}) circle (7pt);
\filldraw[red]   (4.0, 4*sin{60}) circle (7pt);
\filldraw[green] (7.5, 9*sin{60}) circle (7pt);
\filldraw[green] (9.0, 8*sin{60}) circle (7pt);
\filldraw[green] (4.5, 7*sin{60}) circle (7pt);
\filldraw[green] (4.5, 5*sin{60}) circle (7pt);
\filldraw[green] (9.0, 4*sin{60}) circle (7pt);
\filldraw[green] (7.5, 3*sin{60}) circle (7pt);
\end{tikzpicture}
\end{subfigure}
\hfill
\begin{subfigure}[b]{0.19\textwidth}
\caption{\label{fig:5nn}}
\begin{tikzpicture}[scale=0.42]
\clip (2,1) rectangle (10,9);
\draw[line width=3mm,yellow,draw opacity=.5] 
(2,13*sin{60})--(10,8*sin{60})
(2,11*sin{60})--(10,6*sin{60})
(2, 9*sin{60})--(10,4*sin{60})
(2, 7*sin{60})--(10,2*sin{60})
(2, 5*sin{60})--(10,0*sin{60});
  \foreach \i in {0,...,3} 
  \foreach \j in {0,...,5} {
  \foreach \a in {0,120,-120} \draw (3*\i,2*sin{60}*\j) -- +(\a:1);
  \foreach \a in {0,120,-120} \draw (3*\i+3*cos{60},2*sin{60}*\j+sin{60}) -- +(\a:1);}
\draw[ultra thick] (6, 6*sin{60}) circle (3);
\draw[very thick] (6.0, 6*sin{60}) circle (7pt);
\filldraw (9.0, 6*sin{60}) circle (7pt);
\filldraw (3.0, 6*sin{60}) circle (7pt);
\filldraw (7.5, 9*sin{60}) circle (7pt);
\filldraw (7.5, 3*sin{60}) circle (7pt);
\filldraw (4.5, 3*sin{60}) circle (7pt);
\filldraw (4.5, 9*sin{60}) circle (7pt);

\draw[very thick,dotted,blue] (3.0, 6*sin{60})--(4.5, 9*sin{60})-- (7.5, 9*sin{60})--(9.0, 6*sin{60})--(7.5, 3*sin{60})-- (4.5, 3*sin{60})--(3.0, 6*sin{60});

\draw[very thick,dotted,blue]
(6  , 6*sin{60})--(9.0, 6*sin{60})
(6  , 6*sin{60})--(3.0, 6*sin{60})
(6  , 6*sin{60})--(7.5, 9*sin{60}) 
(6  , 6*sin{60})--(7.5, 3*sin{60})
(6  , 6*sin{60})--(4.5, 3*sin{60}) 
(6  , 6*sin{60})--(4.5, 9*sin{60});
\end{tikzpicture}
\end{subfigure}
\caption{\label{fig:hc-basic-neighbours}Subsequent coordination zones on honeycomb lattice:
(a) \textsc{hc}-1, $r^2=1$, $z=3$,
(b) \textsc{hc}-2, $r^2=3$, $z=6$,
(c) \textsc{hc}-3, $r^2=4$, $z=3$,
(d) \textsc{hc}-4, $r^2=27/4$, $z=6$,
(e) \textsc{hc}-5, $r^2=9$, $z=6$.
(a) The lattice constant $a=1$ for \textsc{hc}-1. Lattices (b) \textsc{hc}-2 and (e) \textsc{hc}-5 are equivalent of triangular lattice \textsc{tr}-1 (marked with dotted blue line) with enlarged lattice constant $a=\sqrt 3$ and $3$, respectively.
Lattice (c) \textsc{hc}-3 is equivalent of \textsc{hc}-1 with twice larger lattice constant (marked with dotted red line).
For (a) \textsc{hc}-1, (c) \textsc{hc}-3, (d) \textsc{hc}-4 basic neighborhoods, sites with odd and even labels (marked as red and green open circles) require different implementation of neighbouring sites (marked in red and green solid circles) identification (implemented with the same colors in \Cref{fig:hc-computerized})}
\end{figure*}
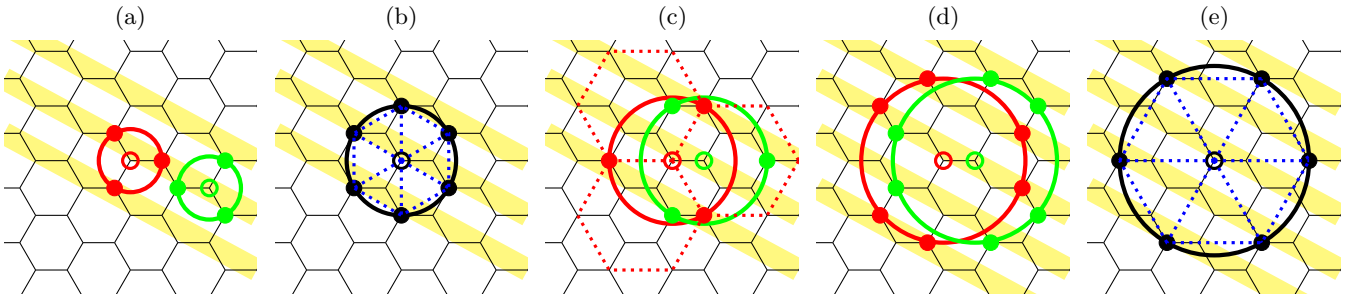

In contrast, noncompact neighborhoods (with holes) destroy dependencies \eqref{eq:pc-discs}, \eqref{eq:pc-finite-z}, \eqref{eq:pc-exp} due to the degeneracy of the percolation threshold (several values of $p_c$ corresponding to the same number $z$ of sites in the neighborhoods). 
This degeneracy is observed for square \cite{Majewski2007} and triangular \cite{2102.10066} lattices.
Here, we show that this degeneracy is also present for the honeycomb lattice, for neighborhoods containing sites up to the fifth coordination zone.
These neighborhoods (see \Cref{fig:hc-neighbours} in \Cref{app:hc-neighbours}) are combined from five basic neighborhoods presented in \Cref{fig:hc-basic-neighbours}.
The lattice names follow the convention proposed in Ref.~\onlinecite{2010.02895} reflecting the lattice topology (here \textsc{hc}, i.e., honeycomb lattice) and a numerical string specifying the coordination zones $i$, where the sites constituting the neighborhood come from.

To avoid the above-mentioned degeneracy, the weighted squared distance $r_i^2$  of  $z_i$  sites in the given neighborhood that belong to the $i$-th coordination zone
\begin{equation} \label{eq:xi} 
\xi=\sum_i z_i r_i^2/i
\end{equation}
has been proposed in Ref.~\onlinecite{2102.10066}.
For a triangular lattice, this index $\xi$ allows differentiation among various neighborhoods and the association of the percolation threshold with its value according to the power law:
\begin{equation} \label{eq:fitpcvsxi} 
p_c\propto\xi^{-\gamma}
\end{equation}
with $\gamma\approx 0.710(19)$ \cite{2102.10066}.

Here, {\em i}) we estimate the site percolation thresholds $p_c$ for the honeycomb lattice with complex neighborhoods containing sites up to the 5-th coordination zone. The second aim of this paper is to check {\em ii}) if the index $\xi$ \eqref{eq:xi} suppresses $p_c(z)$ degeneracy and {\em iii}) if the power law dependence \eqref{eq:fitpcvsxi} proposed for complex neighborhoods and the triangular lattice also holds for the honeycomb lattice.

The paper is organized as follows. 
In \Cref{sec:computations} we describe the details of the calculation of the percolation thresholds and the computer implementation of the proposed methodology.
\Cref{sec:results} contains the results obtained, that is, the percolation thresholds for 31 neighborhoods. 
\Cref{sec:conclusions} is devoted to a discussion of the results and to present the possible directions of further studies.
Finally, three appendices provide: numerical procedures applied here (\Cref{app:functions}); shapes of neighborhoods considered here (\Cref{app:hc-neighbours}); and figures utilized for estimation of percolation thresholds (\Cref{app:PmaxLbetanuvsp}).

\begin{figure}[htbp]
\begin{subfigure}[b]{0.39\textwidth}
\caption{\label{fig:Smax-2}}
\includegraphics[width=0.96\textwidth]{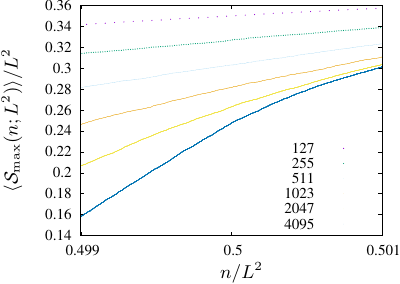}
\end{subfigure}
\begin{subfigure}[b]{0.39\textwidth}
\caption{\label{fig:Pmax-2}}
\includegraphics[width=0.96\textwidth]{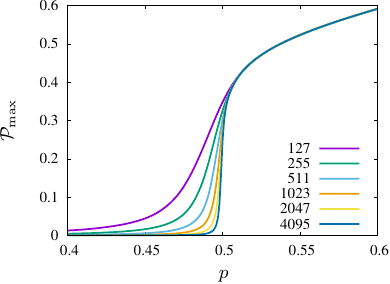}
\end{subfigure}
\begin{subfigure}[b]{0.39\textwidth}
\caption{\label{fig:PmaxLbetanu-2}}
\includegraphics[width=0.96\textwidth]{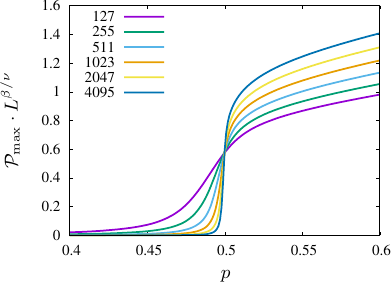}
\end{subfigure}
\caption{\label{fig:PmaxLbetanuvsp-steps}Subsequent stages of the percolation threshold $p_c$ estimations.
(a) $\langle\mathcal{S}_{\max}(n;L^2)\rangle$ (normalized to the system size $L^2$) vs. number of occupied sites $n$ (normalized to the system size $L^2$).
(b) Probability $\mathcal{P}_{\max}$ vs. occupation probability $p$.
(c) Rescaled probability $\mathcal{P}_{\max}\cdot L^{\beta/\nu}$ vs. occupation probability $p$.
The example bases on the results for \textsc{hc}-2 neighborhood and  $R=10^4$ with $\Delta p=0.001$
}
\end{figure}

\section{\label{sec:computations}Computations}

To evaluate the percolation threshold $p_c$ we rely on the finite-size scaling hypothesis \cite[p. 17]{bookDS}, \cite{Finite-Size_Scaling_Theory_1990}, \cite[p. 77]{Guide_to_Monte_Carlo_Simulations_2005}.
According to this hypothesis, near a geometrical phase transition, many quantities $\mathcal A$ obey a scaling relation
\begin{equation}
\label{eq:fss}
\mathcal A(p;L) = L^{-x} \mathcal F\left((p-p_c)L^{1/\nu}\right),
\end{equation}
where $p$ is an occupation probability, $L$ is the linear size of the system, $x$ and $\nu$ are characteristic exponents and $\mathcal F$ is the universal scaling function.
For $p=p_c$, the product $\mathcal A(p;L) L^{x}=\mathcal F(0)$---independently of $L$.
To find $p_c$ we need to plot $L^{x}\mathcal A(p;L)$ for various linear sizes of the system $L$ and the common point of intersection of all these curves predict values of $p_c$.
As an observable $\mathcal A$ we choose the probability $\mathcal P_{\max}$, that a randomly selected site belongs to the largest cluster of occupied sites.
For the two-dimensional percolation problem and the quantity $\mathcal A=\mathcal P_{\max}$ the characteristic exponent $x=\frac{5}{36}/\frac{4}{3}=\frac{5}{48}$ \cite[p. 54]{bookDS}.
The probability $\mathcal P_{\max}$ can be geometrically calculated as the average size of the largest cluster of occupied sites $\langle\mathcal{S}_{\max}\rangle$ divided by the total number of sites $N$.
The brackets $\langle\cdots\rangle$ represent the averaging procedure in the $R$ lattice realization.

To estimate the size of the largest cluster of occupied sites $\mathcal{S}_{\max}$ we utilize the Newman--Ziff algorithm \cite{NewmanZiff2001}.
The algorithm allows you to quickly find $\mathcal S_{\max}$ depending on the number $n$ ($1\le n\le N$) of occupied sites.
This immediately leads to the dependence of $\mathcal S_{\max}(p;N)$, with $p=n/N$, and the size of the system $N$ controls the natural separation $\Delta p=1/N$ between the available values of the probabilities of occupation $p$.
These separations are clearly visible in \Cref{fig:Smax-2} for the smallest system size, $N=127^2$.
To overcome the problem of building $\mathcal S_{\max}(p;N)$ for arbitrarily chosen values of $p$, we again use the idea of Newman and Ziff \cite{NewmanZiff2001}:
\begin{equation}
\mathcal S_{\max}(p;N)=\sum_{n=1}^N \langle \mathcal S_{\max}(n;N)\rangle \mathcal B(n;N,p),
\end{equation}
where $\mathcal B(n;N,p)$ are the binomial distribution coefficients
\begin{equation}
\mathcal B(n;N,p)=\binom{N}{n} p^n (1-p)^{(N-n)}. 
\end{equation}
For large enough systems, these coefficients may be successfully approximated by the normal distribution 
\begin{equation}
\mathcal G(n;\mu,\sigma)=\frac{1}{\sqrt{2\pi\sigma^2}} \exp\left( -\frac{(n-\mu)^2}{2\sigma^2}\right), 
\end{equation}
with $\mu=pN$ and $\sigma^2=p(1-p)N$.
An example of such calculations of the probability that an arbitrarily chosen site belongs to the largest cluster  
\begin{equation}
\label{eq:Pmax}
\mathcal P_{\max}(p;N)=\mathcal S_{\max}(p;N)/ L^2
\end{equation} 
based on $R=10^4$ simulations of $\langle\mathcal S_{\max}(n;N)\rangle$ are presented in \Cref{fig:Pmax-2}.
Finally, the rescaled values of $\mathcal P_{\max}L^{5/48}$ versus $p$---scanned with $\Delta p=10^{-3}$---are presented in \Cref{fig:PmaxLbetanu-2}.

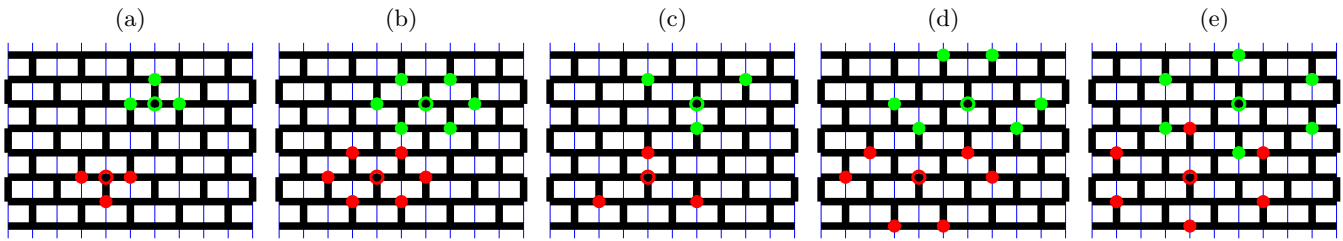
\begin{figure*}[htbp]
\begin{subfigure}[b]{0.19\textwidth}
\caption{\label{fig:comp1nn}}
\begin{tikzpicture}[scale=0.325]
\draw[step=10mm,blue,very thin] (0,1.5) grid (10,9.5);
  \foreach \j in {2,...,9} {\draw[line width=1mm, black]
  (0,\j)--(10,\j);}
  \foreach \i in {0,...,4} 
  \foreach \j in {2,4,6,8} {\draw[line width=1mm,black] (2*\i+1,\j)--(2*\i+1,\j+1);}
  \foreach \i in {0,...,5} 
  \foreach \j in {3,5,7} {\draw[line width=1mm,black]
  (2*\i,\j)--(2*\i,\j+1);}
\draw[very thick,red]   (4,4) circle (7pt);
\filldraw[red]          (3,4) circle (7pt) 
                        (5,4) circle (7pt)
                        (4,3) circle (7pt);
\draw[very thick,green] (6,7) circle (7pt);
\filldraw[green]        (5,7) circle (7pt) 
                        (7,7) circle (7pt)
                        (6,8) circle (7pt);
\end{tikzpicture}
\end{subfigure}
\hfill
\begin{subfigure}[b]{0.19\textwidth}
\caption{\label{fig:comp2nn}}
\begin{tikzpicture}[scale=0.325]
\draw[step=10mm,blue,very thin] (0,1.5) grid (10,9.5);
  \foreach \j in {2,...,9} {\draw[line width=1mm, black]
  (0,\j)--(10,\j);}
  \foreach \i in {0,...,4} 
  \foreach \j in {2,4,6,8} {\draw[line width=1mm,black] (2*\i+1,\j)--(2*\i+1,\j+1);}
  \foreach \i in {0,...,5} 
  \foreach \j in {3,5,7} {\draw[line width=1mm,black]
  (2*\i,\j)--(2*\i,\j+1);}
\draw[very thick,red]   (4,4) circle (7pt);
\filldraw[red]          (2,4) circle (7pt) 
                        (6,4) circle (7pt)
                        (3,3) circle (7pt)
                        (5,3) circle (7pt)
                        (3,5) circle (7pt)
                        (5,5) circle (7pt);
\draw[very thick,green] (6,7) circle (7pt);
\filldraw[green]        (4,7) circle (7pt) 
                        (8,7) circle (7pt)
                        (5,6) circle (7pt)
                        (7,6) circle (7pt)
                        (5,8) circle (7pt)
                        (7,8) circle (7pt);
\end{tikzpicture}
\end{subfigure}
\hfill
\begin{subfigure}[b]{0.19\textwidth}
\caption{\label{fig:comp3nn}}
\begin{tikzpicture}[scale=0.325]
\draw[step=10mm,blue,very thin] (0,1.5) grid (10,9.5);
  \foreach \j in {2,...,9} {\draw[line width=1mm, black]
  (0,\j)--(10,\j);}
  \foreach \i in {0,...,4} 
  \foreach \j in {2,4,6,8} {\draw[line width=1mm,black] (2*\i+1,\j)--(2*\i+1,\j+1);}
  \foreach \i in {0,...,5} 
  \foreach \j in {3,5,7} {\draw[line width=1mm,black]
  (2*\i,\j)--(2*\i,\j+1);}
\draw[very thick,red]   (4,4) circle (7pt);
\filldraw[red]          (2,3) circle (7pt) 
                        (6,3) circle (7pt)
                        (4,5) circle (7pt);
\draw[very thick,green] (6,7) circle (7pt);
\filldraw[green]        (4,8) circle (7pt) 
                        (8,8) circle (7pt)
                        (6,6) circle (7pt);
\end{tikzpicture}
\end{subfigure}
\hfill
\begin{subfigure}[b]{0.19\textwidth}
\caption{\label{fig:comp4nn}}
\begin{tikzpicture}[scale=0.325]
\draw[step=10mm,blue,very thin] (0,1.5) grid (10,9.5);
  \foreach \j in {2,...,9} {\draw[line width=1mm, black]
  (0,\j)--(10,\j);}
  \foreach \i in {0,...,4} 
  \foreach \j in {2,4,6,8} {\draw[line width=1mm,black] (2*\i+1,\j)--(2*\i+1,\j+1);}
  \foreach \i in {0,...,5} 
  \foreach \j in {3,5,7} {\draw[line width=1mm,black]
  (2*\i,\j)--(2*\i,\j+1);}
\draw[very thick,red]   (4,4) circle (7pt);
\filldraw[red]          (1,4) circle (7pt)
                        (7,4) circle (7pt)
                        (2,5) circle (7pt)
                        (6,5) circle (7pt)
                        (3,2) circle (7pt)
                        (5,2) circle (7pt);
\draw[very thick,green] (6,7) circle (7pt);
\filldraw[green]        (3,7) circle (7pt)
                        (9,7) circle (7pt)
                        (5,9) circle (7pt)
                        (7,9) circle (7pt)
                        (4,6) circle (7pt)
                        (8,6) circle (7pt);
\end{tikzpicture}
\end{subfigure}
\hfill
\begin{subfigure}[b]{0.19\textwidth}
\caption{\label{fig:comp5nn}}
\begin{tikzpicture}[scale=0.325]
\draw[step=10mm,blue,very thin] (0,1.5) grid (10,9.5);
  \foreach \j in {2,...,9} {\draw[line width=1mm, black]
  (0,\j)--(10,\j);}
  \foreach \i in {0,...,4} 
  \foreach \j in {2,4,6,8} {\draw[line width=1mm,black] (2*\i+1,\j)--(2*\i+1,\j+1);}
  \foreach \i in {0,...,5} 
  \foreach \j in {3,5,7} {\draw[line width=1mm,black]
  (2*\i,\j)--(2*\i,\j+1);}
\draw[very thick,red]   (4,4) circle (7pt);
\filldraw[red]          (4,2) circle (7pt)
                        (4,6) circle (7pt)
                        (1,5) circle (7pt)
                        (7,5) circle (7pt)
                        (1,3) circle (7pt)
                        (7,3) circle (7pt);
\draw[very thick,green] (6,7) circle (7pt);
\filldraw[green]        (6,5) circle (7pt)
                        (6,9) circle (7pt)
                        (3,8) circle (7pt)
                        (9,8) circle (7pt)
                        (3,6) circle (7pt)
                        (9,6) circle (7pt);
\end{tikzpicture}
\end{subfigure}
\caption{\label{fig:hc-computerized}Honeycomb lattice mapped to a square brick wall lattice~\cite{PhysRevE.60.275}.
(a) \textsc{hc}-1,
(b) \textsc{hc}-2,
(c) \textsc{hc}-3,
(d) \textsc{hc}-4,
(e) \textsc{hc}-5.
Sites at odd and even labels are marked as open red and green circles, respectively.
The color full circles mark sites in the neighbourhood. 
The presented neighborhoods \textsc{hc-}$i$ correspond to subsequent coordination zones $i$.
The horizontal black lines correspond to yellow thick lines on \Cref{fig:hc-basic-neighbours}}
\end{figure*}

Users of the Newman--Ziff algorithm (as implemented in their original paper \cite{NewmanZiff2001}) are responsible for providing {\tt boundaries()} procedures defining the network topology. 
Here, we consider the honeycomb lattice with periodic boundary conditions and complex neighborhoods.
There are 31 such neighborhoods (see \Cref{fig:hc-neighbours} in \Cref{app:hc-neighbours}).
These complex neighborhoods are combined with the basic neighborhoods presented in \Cref{fig:hc-basic-neighbours}. 

The underlying neighborhoods can be mapped into a square brick wall lattice \cite{PhysRevE.60.275}.
For the \textsc{hc-1}, \textsc{hc-3}, \textsc{hc-4} lattices, the implementation of neighboring sites must be separated into odd and even site labels, as the honeycomb lattice is excluded from the set of two-dimensional Bravais lattices \cite[p. 8]{Kittel_2005}.
These computerized versions of lattices with basic neighborhoods (\Cref{fig:hc-basic-neighbours}) mapped on a square brick wall lattice are presented in \Cref{fig:hc-computerized}. 
The {\tt boundaries()} functions necessary for the Newman--Ziff algorithm are presented in \Cref{app:functions}.

\section{\label{sec:results}Results}

\begin{table}[!h]
\caption{\label{tab:pc-5}Estimated values of percolation thresholds $p_c$ for various complex neighborhoods. 
The lattice name encodes the involved coordination zones ($i\le 5$), to which sites in the neighbourhood belong.
Also the coordination number $z$ and the indexes $\xi$ \eqref{eq:xi} and $\zeta$ \eqref{eq:zeta} are presented}
\begin{ruledtabular}
\begin{tabular}{lrrrl}
lattice                 & $z$ & $\xi$ & $\zeta$ & $p_c$\\
\hline 
\hline 
\textsc{hc}-1           &  3 &  3 &  3 &0.6970\footnote{0.697043(3)~\cite{PhysRevE.60.275},
                                        0.700~\cite{Iribarne1999},
                                        0.6962(6)~\cite{Djordjevic_1982}, 
                                        0.697040~\cite{Jacobsen_2014},
                                        0.697040230(5)~\cite{Jacobsen_2014}, 
                                        0.6970402(1)~\cite{PhysRevE.78.031136}}\\
\hline 
\textsc{hc}-2\footnote{equivalent \textsc{tr}-1} &  6 &  9 & 10.3923 & 0.5000\footnote{$\frac{1}{2}$ \cite[p. 17]{bookDS}}\\
\textsc{hc}-1,2         &  9 & 12 & 13.3923 & 0.3630\footnote{0.359~\cite{Iribarne1999}}\\
\hline 
\textsc{hc}-3\footnote{equivalent \textsc{hc}-1} & 3 &  4 & 6 & 0.6970\footnotemark[1]\\
\textsc{hc}-1,3       	&  6 &  7 & 9 & 0.4132\\
\textsc{hc}-2,3       	&  9 & 13 & 16.3923 & 0.3139\\
\textsc{hc}-1,2,3       & 12 & 16 & 19.3923 & 0.3030\footnote{0.300~\cite{Dalton_1964}, 0.302960~\cite{PhysRevE.105.024105}}\\
\hline 
\textsc{hc}-4           &  6 & 10.125 &15.5885& 0.3154\\
\textsc{hc}-1,4       	&  9 & 13.125 &18.5885& 0.2704\\
\textsc{hc}-2,4       	& 12 & 19.125 &25.9808& 0.2374\\
\textsc{hc}-3,4         &  9 & 14.125 &21.5885& 0.2556\\
\textsc{hc}-1,2,4     	& 15 & 22.125 &28.9808& 0.2278\\
\textsc{hc}-1,3,4     	& 12 & 17.125 &24.5885& 0.2364\\
\textsc{hc}-2,3,4     	& 15 & 23.125 &31.9808& 0.2161\\
\textsc{hc}-1,2,3,4   	& 18 & 26.125 &34.9808& 0.2113\footnote{0.210~\cite{Iribarne1999}}\\
\hline 
\textsc{hc}-5\footnotemark[2] &  6    & 10.8 &18& 0.5000\footnotemark[3]\\
\textsc{hc}-1,5       	&  9 &  13.8  &21     & 0.2654\\
\textsc{hc}-2,5         & 12 &  19.8  &28.3923& 0.2903\\
\textsc{hc}-3,5       	&  9 &  14.8  &24     & 0.2560\\
\textsc{hc}-4,5         & 12 & 20.925 &33.5885& 0.2014\\
\textsc{hc}-1,2,5     	& 15 & 22.8   &31.3923& 0.2147\\
\textsc{hc}-1,3,5     	& 12 & 17.8   &27     & 0.2290\\
\textsc{hc}-1,4,5     	& 15 & 23.925 &36.5885& 0.1913\\
\textsc{hc}-2,3,5     	& 15 & 23.8   &34.3923& 0.2043\\
\textsc{hc}-2,4,5     	& 18 & 29.925 &43.9808& 0.1752\\
\textsc{hc}-3,4,5     	& 15 & 24.925 &39.5885& 0.1863\\
\textsc{hc}-1,2,3,5   	& 18 & 26.8   &37.3923& 0.1973\\
\textsc{hc}-1,2,4,5   	& 21 & 32.925 &46.9808& 0.1720\\	
\textsc{hc}-1,3,4,5   	& 18 & 27.925 &42.5885& 0.1795\\	
\textsc{hc}-2,3,4,5   	& 21 & 33.925 &49.9808& 0.1673\\
\textsc{hc}-1,2,3,4,5 	& 24 & 36.925 &52.9808& 0.1655\footnote{0.164~\cite{Iribarne1999}}\\
\end{tabular}
\end{ruledtabular}
\end{table}

In \Cref{fig:PmaxLbetanuvsp-various} (in \Cref{app:PmaxLbetanuvsp}) the dependencies $\mathcal P_{\max} L^{5/48}$ on the probability of occupation $p$ obtained by the procedure described in \Cref{sec:computations} (see \Cref{fig:PmaxLbetanuvsp-steps}) are presented.
Data are based on $\langle\mathcal S_{\max}(n,L^2)\rangle$ simulated with the Newman--Ziff algorithm for $N=L^2$ sites and $L=127$, 255, 511, 1023, 2047, and 4095 averaged over $R=10^4$ simulations with $\Delta p=10^{-4}$ data separation.
The percolation thresholds $p_c$ predicted by a common point of six curves for various $L$ with an accuracy given by an assumed data separation constant $\Delta p=10^{-4}$ are collected in \Cref{tab:pc-5}.
The only exception is \textsc{hc-3}, where the curves for various $L$ do not intercept each other at a single point (see \Cref{fig:PmaxLbetanuvsp-3}).
Fortunately, $p_c(\textsc{hc-3})=p_c(\textsc{hc-1})$ as the \textsc{hc-3} lattice is equivalent to \textsc{hc-1} due to the symmetry of the neighborhood analysis (see \Cref{fig:3nn}).
 
\section{\label{sec:conclusions}Conclusions}

\begin{figure}
\begin{subfigure}[b]{0.40\textwidth}
\caption{\label{fig:hc-pc_vs_z}}
\includegraphics[width=.880\textwidth]{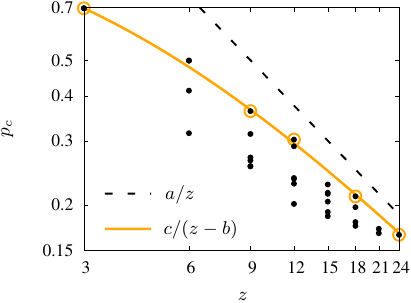}
\end{subfigure}
\begin{subfigure}[b]{0.40\textwidth}
\caption{\label{fig:hc-pc_vs_xi}}
\includegraphics[width=.880\textwidth]{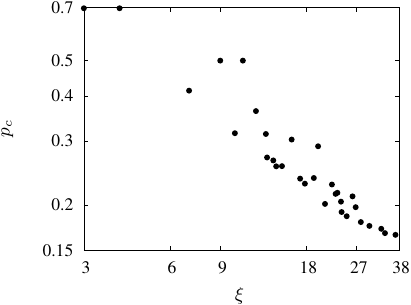}
\end{subfigure}
\begin{subfigure}[b]{0.40\textwidth}
\caption{\label{fig:hc-pc_vs_zeta}}
\includegraphics[width=.880\textwidth]{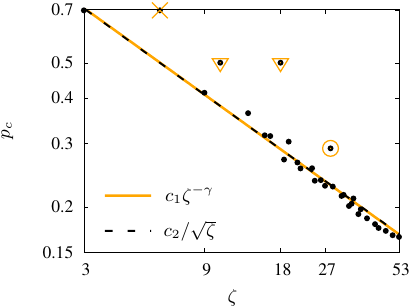}
\end{subfigure}
\caption{\label{fig:pcvsz}Percolation thresholds $p_c$ for complex neighborhoods in the honeycomb lattice. 
(a) Dependence of degenerated values of $p_c$ on the total number $z$ of sites in the neighborhood. 
The dependence \eqref{eq:pc-discs} is also presented by a dashed line.
(b) Dependence of $p_c$ vs. index $\xi$ \eqref{eq:xi} for complex neighborhoods.
The $p_c$ values for equivalents of \textsc{hc-1} and \textsc{hc-2} are marked with crosses.
(c) Dependence of $p_c$ vs. index $\zeta$ \eqref{eq:zeta} for complex neighborhoods.
The $p_c$ values for equivalents of \textsc{hc-1} (\textsc{hc-3}, marked with cross) and \textsc{tr-1} (\textsc{hc-2} and \textsc{hc-5}, marked with triangles) neighborhoods are excluded from the fitting.
The percolation threshold for \textsc{hc-2,5} (marked with an open circle) is also excluded from a fitting procedure}
\end{figure}

For complex and noncompact neighborhoods, the total number $z$ of sites in the neighborhood is insufficient to differentiate between neighborhoods in terms of the percolation thresholds associated with this type of neighborhood (see \Cref{fig:hc-pc_vs_z}). 
The limiting case [of discs, according to \Cref{eq:pc-discs}] with $a=4.512$ \cite{PhysRevE.105.024105} and its finite-$z$ corrections [\Cref{eq:pc-finite-z}] are presented there with black dashed and orange solid lines, respectively.
In the latter case, the least squares fit gives $c=4.630(75)$ and $b=3.64(13)$. 
The fitting procedure were performed via five points corresponding to the compact neighborhoods (\textsc{hc-1}, \textsc{hc-12}, \textsc{hc-123}, \textsc{hc-1234}, \textsc{hc-12345}) marked by the open circles.

The degeneracy of $p_c(z)$ mentioned above can be solved by introducing the index $\xi$ [\Cref{eq:xi}] as proposed in Ref.~\onlinecite{2102.10066}.
In fact, for the triangular lattice and complex neighborhoods, the various neighborhoods are characterized by various values of the $\xi$ index \cite{2102.10066}.
The utility of the $\xi$ index to distinguish between various neighborhoods is also the same for the honeycomb lattice (see \Cref{fig:hc-pc_vs_xi}).
Unfortunately, the power law dependency of $p_c$ on $\xi$ does not apply to the honeycomb lattice (see \Cref{fig:hc-pc_vs_xi}).
Thus, we propose yet another heuristic index  
\begin{equation} \label{eq:zeta} 
\zeta=\sum_i z_i r_i,
\end{equation}
which seems to be much more appropriate here and which gives a much better (but not perfect) power fit 
\begin{equation} \label{eq:pcvszeta} 
p_c=c_1\zeta^{-\gamma}
\end{equation}
with $\gamma\approx\frac{1}{2}$ (the fit according to the least squares method gives $\gamma=0.4981(90)$).
The $p_c$ values for the equivalents of the \textsc{hc-1} (\textsc{hc-3}, marked with a cross) and \textsc{tr-1} (\textsc{hc-2} and \textsc{hc-5}, marked with triangles) and \textsc{hc-2,5} (marked with an open circle) neighborhoods are excluded from the fitting.
Basing on the obtained value of $\gamma$ we have also checked the inverse square root dependence of the percolation threshold on the $\zeta$ index
\begin{equation} \label{eq:pcvssqrtzeta} 
p_c=c_2/\sqrt{\zeta}
\end{equation}
with a fitted value of constant $c_2=1.2251(99)$.
The dependence of the percolation threshold $p_c$ on the newly proposed index $\zeta$ and the fits according to \Cref{eq:pcvszeta,eq:pcvssqrtzeta} are presented in \Cref{fig:hc-pc_vs_zeta} with solid orange and black dashed lines, respectively.

The values of $\xi$ and $\zeta$ are also presented in \Cref{tab:pc-5}. 

In conclusion, basing on the Newman--Ziff algorithm and finite-size scaling analysis we have calculated the random site percolation thresholds $p_c$ for complex neighbourhoods on honeycomb lattice.
We propose a scalar quantity $\zeta$ which may be helpful for differentiating among various neighbourhoods.
The quantity is based on sites number $z_i$ and sites distances $r_i$ to the central site in the sites $i$-th coordination zone.
The dependency of $p_c$ on this newly proposed index $\zeta$ follows roughly a inverse square root dependence $p_c\propto 1/\sqrt\zeta$.

\bibliography{percolation,basics,km}

\begin{thebibliography}{35}%
\makeatletter
\providecommand \@ifxundefined [1]{%
 \@ifx{#1\undefined}
}%
\providecommand \@ifnum [1]{%
 \ifnum #1\expandafter \@firstoftwo
 \else \expandafter \@secondoftwo
 \fi
}%
\providecommand \@ifx [1]{%
 \ifx #1\expandafter \@firstoftwo
 \else \expandafter \@secondoftwo
 \fi
}%
\providecommand \natexlab [1]{#1}%
\providecommand \enquote  [1]{``#1''}%
\providecommand \bibnamefont  [1]{#1}%
\providecommand \bibfnamefont [1]{#1}%
\providecommand \citenamefont [1]{#1}%
\providecommand \href@noop [0]{\@secondoftwo}%
\providecommand \href [0]{\begingroup \@sanitize@url \@href}%
\providecommand \@href[1]{\@@startlink{#1}\@@href}%
\providecommand \@@href[1]{\endgroup#1\@@endlink}%
\providecommand \@sanitize@url [0]{\catcode `\\12\catcode `\$12\catcode
  `\&12\catcode `\#12\catcode `\^12\catcode `\_12\catcode `\%12\relax}%
\providecommand \@@startlink[1]{}%
\providecommand \@@endlink[0]{}%
\providecommand \url  [0]{\begingroup\@sanitize@url \@url }%
\providecommand \@url [1]{\endgroup\@href {#1}{\urlprefix }}%
\providecommand \urlprefix  [0]{URL }%
\providecommand \Eprint [0]{\href }%
\providecommand \doibase [0]{http://dx.doi.org/}%
\providecommand \selectlanguage [0]{\@gobble}%
\providecommand \bibinfo  [0]{\@secondoftwo}%
\providecommand \bibfield  [0]{\@secondoftwo}%
\providecommand \translation [1]{[#1]}%
\providecommand \BibitemOpen [0]{}%
\providecommand \bibitemStop [0]{}%
\providecommand \bibitemNoStop [0]{.\EOS\space}%
\providecommand \EOS [0]{\spacefactor3000\relax}%
\providecommand \BibitemShut  [1]{\csname bibitem#1\endcsname}%
\let\auto@bib@innerbib\@empty
\bibitem [{\citenamefont {Broadbent}\ and\ \citenamefont
  {Hammersley}(1957)}]{Broadbent1957}%
  \BibitemOpen
  \bibfield  {author} {\bibinfo {author} {\bibfnamefont {S.~R.}\ \bibnamefont
  {Broadbent}}\ and\ \bibinfo {author} {\bibfnamefont {J.~M.}\ \bibnamefont
  {Hammersley}},\ }\bibfield  {title} {\enquote {\bibinfo {title} {Percolation
  processes: I. {C}rystals and mazes},}\ }\href {\doibase
  10.1017/S0305004100032680} {\bibfield  {journal} {\bibinfo  {journal}
  {Mathematical Proceedings of the Cambridge Philosophical Society}\ }\textbf
  {\bibinfo {volume} {53}},\ \bibinfo {pages} {629--641} (\bibinfo {year}
  {1957})}\BibitemShut {NoStop}%
\bibitem [{\citenamefont {Hammersley}(1957)}]{Hammersley1957}%
  \BibitemOpen
  \bibfield  {author} {\bibinfo {author} {\bibfnamefont {J.~M.}\ \bibnamefont
  {Hammersley}},\ }\bibfield  {title} {\enquote {\bibinfo {title} {Percolation
  processes: {II}. {T}he connective constant},}\ }\href {\doibase
  10.1017/S0305004100032692} {\bibfield  {journal} {\bibinfo  {journal}
  {Mathematical Proceedings of the Cambridge Philosophical Society}\ }\textbf
  {\bibinfo {volume} {53}},\ \bibinfo {pages} {642--645} (\bibinfo {year}
  {1957})}\BibitemShut {NoStop}%
\bibitem [{\citenamefont {Stauffer}\ and\ \citenamefont
  {Aharony}(1994)}]{bookDS}%
  \BibitemOpen
  \bibfield  {author} {\bibinfo {author} {\bibfnamefont {D.}~\bibnamefont
  {Stauffer}}\ and\ \bibinfo {author} {\bibfnamefont {A.}~\bibnamefont
  {Aharony}},\ }\href {\doibase 10.1201/9781315274386} {\emph {\bibinfo {title}
  {Introduction to Percolation Theory}}},\ \bibinfo {edition} {2nd}\ ed.\
  (\bibinfo  {publisher} {Taylor and Francis},\ \bibinfo {address} {London},\
  \bibinfo {year} {1994})\BibitemShut {NoStop}%
\bibitem [{\citenamefont {Wierman}(2014)}]{Wierman2014}%
  \BibitemOpen
  \bibfield  {author} {\bibinfo {author} {\bibfnamefont {J.}~\bibnamefont
  {Wierman}},\ }\enquote {\bibinfo {title} {Percolation theory},}\ in\ \href
  {\doibase 10.1002/9781118445112.stat02317} {\emph {\bibinfo {booktitle}
  {Wiley StatsRef: Statistics Reference Online}}}\ (\bibinfo  {publisher}
  {American Cancer Society},\ \bibinfo {year} {2014})\ pp.\ \bibinfo {pages}
  {1--9}\BibitemShut {NoStop}%
\bibitem [{wik(2020)}]{wiki.Percolation_threhold}%
  \BibitemOpen
  \href {https://en.wikipedia.org/wiki/Percolation_threshold} {\enquote
  {\bibinfo {title} {en.wikipedia.org/wiki/percolation\_threshold},}\ }
  (\bibinfo {year} {2020})\BibitemShut {NoStop}%
\bibitem [{\citenamefont {Li}\ \emph {et~al.}(2021)\citenamefont {Li},
  \citenamefont {Liu}, \citenamefont {L\"u}, \citenamefont {Hu}, \citenamefont
  {Xu},\ and\ \citenamefont {Zhang}}]{Li_2021}%
  \BibitemOpen
  \bibfield  {author} {\bibinfo {author} {\bibfnamefont {M.}~\bibnamefont
  {Li}}, \bibinfo {author} {\bibfnamefont {R.-R.}\ \bibnamefont {Liu}},
  \bibinfo {author} {\bibfnamefont {L.}~\bibnamefont {L\"u}}, \bibinfo {author}
  {\bibfnamefont {M.-B.}\ \bibnamefont {Hu}}, \bibinfo {author} {\bibfnamefont
  {S.}~\bibnamefont {Xu}}, \ and\ \bibinfo {author} {\bibfnamefont {Y.-C.}\
  \bibnamefont {Zhang}},\ }\bibfield  {title} {\enquote {\bibinfo {title}
  {Percolation on complex networks: {T}heory and application},}\ }\href
  {\doibase 10.1016/j.physrep.2020.12.003} {\bibfield  {journal} {\bibinfo
  {journal} {Physics Reports}\ ,\ \bibinfo {pages} {1--68}} (\bibinfo {year}
  {2021})}\BibitemShut {NoStop}%
\bibitem [{\citenamefont {Saberi}(2015)}]{Saberi2015}%
  \BibitemOpen
  \bibfield  {author} {\bibinfo {author} {\bibfnamefont {A.~A.}\ \bibnamefont
  {Saberi}},\ }\bibfield  {title} {\enquote {\bibinfo {title} {Recent advances
  in percolation theory and its applications},}\ }\href {\doibase
  10.1016/j.physrep.2015.03.003} {\bibfield  {journal} {\bibinfo  {journal}
  {Physics Reports}\ }\textbf {\bibinfo {volume} {578}},\ \bibinfo {pages}
  {1--32} (\bibinfo {year} {2015})}\BibitemShut {NoStop}%
\bibitem [{\citenamefont {Dean}(1963)}]{Dean_1963}%
  \BibitemOpen
  \bibfield  {author} {\bibinfo {author} {\bibfnamefont {P.}~\bibnamefont
  {Dean}},\ }\bibfield  {title} {\enquote {\bibinfo {title} {A new {M}onte
  {C}arlo method for percolation problems on a lattice},}\ }\href {\doibase
  10.1017/S0305004100037026} {\bibfield  {journal} {\bibinfo  {journal}
  {Mathematical Proceedings of the Cambridge Philosophical Society}\ }\textbf
  {\bibinfo {volume} {59}},\ \bibinfo {pages} {397--410} (\bibinfo {year}
  {1963})}\BibitemShut {NoStop}%
\bibitem [{\citenamefont {Dean}\ and\ \citenamefont
  {Bird}(1967)}]{Dean_Bird_1967}%
  \BibitemOpen
  \bibfield  {author} {\bibinfo {author} {\bibfnamefont {P.}~\bibnamefont
  {Dean}}\ and\ \bibinfo {author} {\bibfnamefont {N.~F.}\ \bibnamefont
  {Bird}},\ }\bibfield  {title} {\enquote {\bibinfo {title} {Monte {C}arlo
  estimates of critical percolation probabilities},}\ }\href {\doibase
  10.1017/S0305004100041438} {\bibfield  {journal} {\bibinfo  {journal}
  {Mathematical Proceedings of the Cambridge Philosophical Society}\ }\textbf
  {\bibinfo {volume} {63}},\ \bibinfo {pages} {477--479} (\bibinfo {year}
  {1967})}\BibitemShut {NoStop}%
\bibitem [{\citenamefont {Suding}\ and\ \citenamefont
  {Ziff}(1999)}]{PhysRevE.60.275}%
  \BibitemOpen
  \bibfield  {author} {\bibinfo {author} {\bibfnamefont {P.~N.}\ \bibnamefont
  {Suding}}\ and\ \bibinfo {author} {\bibfnamefont {R.~M.}\ \bibnamefont
  {Ziff}},\ }\bibfield  {title} {\enquote {\bibinfo {title} {Site percolation
  thresholds for {A}rchimedean lattices},}\ }\href {\doibase
  10.1103/PhysRevE.60.275} {\bibfield  {journal} {\bibinfo  {journal} {Physical
  Review E}\ }\textbf {\bibinfo {volume} {60}},\ \bibinfo {pages} {275--283}
  (\bibinfo {year} {1999})}\BibitemShut {NoStop}%
\bibitem [{\citenamefont {Dalton}\ \emph {et~al.}(1964)\citenamefont {Dalton},
  \citenamefont {Domb},\ and\ \citenamefont {Sykes}}]{Dalton_1964}%
  \BibitemOpen
  \bibfield  {author} {\bibinfo {author} {\bibfnamefont {N.~W.}\ \bibnamefont
  {Dalton}}, \bibinfo {author} {\bibfnamefont {C.}~\bibnamefont {Domb}}, \ and\
  \bibinfo {author} {\bibfnamefont {M.~F.}\ \bibnamefont {Sykes}},\ }\bibfield
  {title} {\enquote {\bibinfo {title} {Dependence of critical concentration of
  a dilute ferromagnet on the range of interaction},}\ }\href {\doibase
  10.1088/0370-1328/83/3/118} {\bibfield  {journal} {\bibinfo  {journal}
  {Proceedings of the Physical Society}\ }\textbf {\bibinfo {volume} {83}},\
  \bibinfo {pages} {496--498} (\bibinfo {year} {1964})}\BibitemShut {NoStop}%
\bibitem [{\citenamefont {Domb}\ and\ \citenamefont {Dalton}(1966)}]{Domb1966}%
  \BibitemOpen
  \bibfield  {author} {\bibinfo {author} {\bibfnamefont {C.}~\bibnamefont
  {Domb}}\ and\ \bibinfo {author} {\bibfnamefont {N.~W.}\ \bibnamefont
  {Dalton}},\ }\bibfield  {title} {\enquote {\bibinfo {title} {Crystal
  statistics with long-range forces: {I}. {T}he equivalent neighbour model},}\
  }\href {\doibase 10.1088/0370-1328/89/4/311} {\bibfield  {journal} {\bibinfo
  {journal} {Proceedings of the Physical Society}\ }\textbf {\bibinfo {volume}
  {89}},\ \bibinfo {pages} {859--871} (\bibinfo {year} {1966})}\BibitemShut
  {NoStop}%
\bibitem [{\citenamefont {Gouker}\ and\ \citenamefont
  {Family}(1983)}]{Gouker1983}%
  \BibitemOpen
  \bibfield  {author} {\bibinfo {author} {\bibfnamefont {M.}~\bibnamefont
  {Gouker}}\ and\ \bibinfo {author} {\bibfnamefont {F.}~\bibnamefont
  {Family}},\ }\bibfield  {title} {\enquote {\bibinfo {title} {Evidence for
  classical critical behavior in long-range site percolation},}\ }\href
  {\doibase 10.1103/PhysRevB.28.1449} {\bibfield  {journal} {\bibinfo
  {journal} {Physical Review B}\ }\textbf {\bibinfo {volume} {28}},\ \bibinfo
  {pages} {1449--1452} (\bibinfo {year} {1983})}\BibitemShut {NoStop}%
\bibitem [{\citenamefont {Malarz}\ and\ \citenamefont
  {Galam}(2005)}]{Galam2005a}%
  \BibitemOpen
  \bibfield  {author} {\bibinfo {author} {\bibfnamefont {K.}~\bibnamefont
  {Malarz}}\ and\ \bibinfo {author} {\bibfnamefont {S.}~\bibnamefont {Galam}},\
  }\bibfield  {title} {\enquote {\bibinfo {title} {Square-lattice site
  percolation at increasing ranges of neighbor bonds},}\ }\href {\doibase
  10.1103/PhysRevE.71.016125} {\bibfield  {journal} {\bibinfo  {journal}
  {Physical Review E}\ }\textbf {\bibinfo {volume} {71}},\ \bibinfo {pages}
  {016125} (\bibinfo {year} {2005})}\BibitemShut {NoStop}%
\bibitem [{\citenamefont {Galam}\ and\ \citenamefont
  {Malarz}(2005)}]{Galam2005b}%
  \BibitemOpen
  \bibfield  {author} {\bibinfo {author} {\bibfnamefont {S.}~\bibnamefont
  {Galam}}\ and\ \bibinfo {author} {\bibfnamefont {K.}~\bibnamefont {Malarz}},\
  }\bibfield  {title} {\enquote {\bibinfo {title} {Restoring site percolation
  on damaged square lattices},}\ }\href {\doibase 10.1103/PhysRevE.72.027103}
  {\bibfield  {journal} {\bibinfo  {journal} {Physical Review E}\ }\textbf
  {\bibinfo {volume} {72}},\ \bibinfo {pages} {027103} (\bibinfo {year}
  {2005})}\BibitemShut {NoStop}%
\bibitem [{\citenamefont {Majewski}\ and\ \citenamefont
  {Malarz}(2007)}]{Majewski2007}%
  \BibitemOpen
  \bibfield  {author} {\bibinfo {author} {\bibfnamefont {M.}~\bibnamefont
  {Majewski}}\ and\ \bibinfo {author} {\bibfnamefont {K.}~\bibnamefont
  {Malarz}},\ }\bibfield  {title} {\enquote {\bibinfo {title} {Square lattice
  site percolation thresholds for complex neighbourhoods},}\ }\href
  {http://www.actaphys.uj.edu.pl/fulltext?series=Reg&vol=38&page=2191}
  {\bibfield  {journal} {\bibinfo  {journal} {Acta Physica Polonica B}\
  }\textbf {\bibinfo {volume} {38}},\ \bibinfo {pages} {2191--2199} (\bibinfo
  {year} {2007})}\BibitemShut {NoStop}%
\bibitem [{\citenamefont {Xun}\ \emph {et~al.}(2021)\citenamefont {Xun},
  \citenamefont {Hao},\ and\ \citenamefont {Ziff}}]{2010.02895}%
  \BibitemOpen
  \bibfield  {author} {\bibinfo {author} {\bibfnamefont {Z.}~\bibnamefont
  {Xun}}, \bibinfo {author} {\bibfnamefont {D.}~\bibnamefont {Hao}}, \ and\
  \bibinfo {author} {\bibfnamefont {R.~M.}\ \bibnamefont {Ziff}},\ }\bibfield
  {title} {\enquote {\bibinfo {title} {Site percolation on square and simple
  cubic lattices with extended neighborhoods and their continuum limit},}\
  }\href {\doibase 10.1103/PhysRevE.103.022126} {\bibfield  {journal} {\bibinfo
   {journal} {Physical Review E}\ }\textbf {\bibinfo {volume} {103}},\ \bibinfo
  {pages} {022126} (\bibinfo {year} {2021})}\BibitemShut {NoStop}%
\bibitem [{\citenamefont {d'Iribarne}\ \emph {et~al.}(1999)\citenamefont
  {d'Iribarne}, \citenamefont {Rasigni},\ and\ \citenamefont
  {Rasigni}}]{Iribarne1999}%
  \BibitemOpen
  \bibfield  {author} {\bibinfo {author} {\bibfnamefont {C.}~\bibnamefont
  {d'Iribarne}}, \bibinfo {author} {\bibfnamefont {M.}~\bibnamefont {Rasigni}},
  \ and\ \bibinfo {author} {\bibfnamefont {G.}~\bibnamefont {Rasigni}},\
  }\bibfield  {title} {\enquote {\bibinfo {title} {From lattice long-range
  percolation to the continuum one},}\ }\href {\doibase
  10.1016/S0375-9601(99)00585-X} {\bibfield  {journal} {\bibinfo  {journal}
  {Physics Letters A}\ }\textbf {\bibinfo {volume} {263}},\ \bibinfo {pages}
  {65--69} (\bibinfo {year} {1999})}\BibitemShut {NoStop}%
\bibitem [{\citenamefont {Malarz}(2020)}]{2006.15621}%
  \BibitemOpen
  \bibfield  {author} {\bibinfo {author} {\bibfnamefont {K.}~\bibnamefont
  {Malarz}},\ }\bibfield  {title} {\enquote {\bibinfo {title} {Site percolation
  thresholds on triangular lattice with complex neighborhoods},}\ }\href
  {\doibase 10.1063/5.0022336} {\bibfield  {journal} {\bibinfo  {journal}
  {Chaos}\ }\textbf {\bibinfo {volume} {30}},\ \bibinfo {pages} {123123}
  (\bibinfo {year} {2020})}\BibitemShut {NoStop}%
\bibitem [{\citenamefont {Malarz}(2021)}]{2102.10066}%
  \BibitemOpen
  \bibfield  {author} {\bibinfo {author} {\bibfnamefont {K.}~\bibnamefont
  {Malarz}},\ }\bibfield  {title} {\enquote {\bibinfo {title} {Percolation
  thresholds on triangular lattice for neighbourhoods containing sites up-to
  the fifth coordination zone},}\ }\href {\doibase 10.1103/PhysRevE.103.052107}
  {\bibfield  {journal} {\bibinfo  {journal} {Physical Review E}\ }\textbf
  {\bibinfo {volume} {103}},\ \bibinfo {pages} {052107} (\bibinfo {year}
  {2021})}\BibitemShut {NoStop}%
\bibitem [{\citenamefont {Xun}\ \emph {et~al.}(2022)\citenamefont {Xun},
  \citenamefont {Hao},\ and\ \citenamefont {Ziff}}]{PhysRevE.105.024105}%
  \BibitemOpen
  \bibfield  {author} {\bibinfo {author} {\bibfnamefont {Z.}~\bibnamefont
  {Xun}}, \bibinfo {author} {\bibfnamefont {D.}~\bibnamefont {Hao}}, \ and\
  \bibinfo {author} {\bibfnamefont {R.~M.}\ \bibnamefont {Ziff}},\ }\bibfield
  {title} {\enquote {\bibinfo {title} {Site and bond percolation thresholds on
  regular lattices with compact extended-range neighborhoods in two and three
  dimensions},}\ }\href {\doibase 10.1103/PhysRevE.105.024105} {\bibfield
  {journal} {\bibinfo  {journal} {Physical Review E}\ }\textbf {\bibinfo
  {volume} {105}},\ \bibinfo {pages} {024105} (\bibinfo {year}
  {2022})}\BibitemShut {NoStop}%
\bibitem [{\citenamefont {Kurzawski}\ and\ \citenamefont
  {Malarz}(2012)}]{Kurzawski2012}%
  \BibitemOpen
  \bibfield  {author} {\bibinfo {author} {\bibfnamefont {{\L}.}~\bibnamefont
  {Kurzawski}}\ and\ \bibinfo {author} {\bibfnamefont {K.}~\bibnamefont
  {Malarz}},\ }\bibfield  {title} {\enquote {\bibinfo {title} {Simple cubic
  random-site percolation thresholds for complex neighbourhoods},}\ }\href
  {\doibase 10.1016/S0034-4877(12)60036-6} {\bibfield  {journal} {\bibinfo
  {journal} {Reports on Mathematical Physics}\ }\textbf {\bibinfo {volume}
  {70}},\ \bibinfo {pages} {163--169} (\bibinfo {year} {2012})}\BibitemShut
  {NoStop}%
\bibitem [{\citenamefont {Malarz}(2015)}]{Malarz2015}%
  \BibitemOpen
  \bibfield  {author} {\bibinfo {author} {\bibfnamefont {K.}~\bibnamefont
  {Malarz}},\ }\bibfield  {title} {\enquote {\bibinfo {title} {Simple cubic
  random-site percolation thresholds for neighborhoods containing
  fourth-nearest neighbors},}\ }\href {\doibase 10.1103/PhysRevE.91.043301}
  {\bibfield  {journal} {\bibinfo  {journal} {Physical Review E}\ }\textbf
  {\bibinfo {volume} {91}},\ \bibinfo {pages} {043301} (\bibinfo {year}
  {2015})}\BibitemShut {NoStop}%
\bibitem [{\citenamefont {Kotwica}\ \emph {et~al.}(2019)\citenamefont
  {Kotwica}, \citenamefont {Gronek},\ and\ \citenamefont
  {Malarz}}]{1803.09504}%
  \BibitemOpen
  \bibfield  {author} {\bibinfo {author} {\bibfnamefont {M.}~\bibnamefont
  {Kotwica}}, \bibinfo {author} {\bibfnamefont {P.}~\bibnamefont {Gronek}}, \
  and\ \bibinfo {author} {\bibfnamefont {K.}~\bibnamefont {Malarz}},\
  }\bibfield  {title} {\enquote {\bibinfo {title} {Efficient space
  virtualisation for {H}oshen--{K}opelman algorithm},}\ }\href {\doibase
  10.1142/S0129183119500554} {\bibfield  {journal} {\bibinfo  {journal}
  {International Journal of Modern Physics C}\ }\textbf {\bibinfo {volume}
  {30}},\ \bibinfo {pages} {1950055} (\bibinfo {year} {2019})}\BibitemShut
  {NoStop}%
\bibitem [{\citenamefont {Coupette}\ and\ \citenamefont
  {Schilling}(2022)}]{PhysRevE.105.044108}%
  \BibitemOpen
  \bibfield  {author} {\bibinfo {author} {\bibfnamefont {F.}~\bibnamefont
  {Coupette}}\ and\ \bibinfo {author} {\bibfnamefont {T.}~\bibnamefont
  {Schilling}},\ }\bibfield  {title} {\enquote {\bibinfo {title} {Exactly
  solvable percolation problems},}\ }\href {\doibase
  10.1103/PhysRevE.105.044108} {\bibfield  {journal} {\bibinfo  {journal}
  {Physical Review E}\ }\textbf {\bibinfo {volume} {105}},\ \bibinfo {pages}
  {044108} (\bibinfo {year} {2022})}\BibitemShut {NoStop}%
\bibitem [{\citenamefont {Xu}\ \emph {et~al.}(2021)\citenamefont {Xu},
  \citenamefont {Wang}, \citenamefont {Hu},\ and\ \citenamefont
  {Deng}}]{PhysRevE.103.022127}%
  \BibitemOpen
  \bibfield  {author} {\bibinfo {author} {\bibfnamefont {W.}~\bibnamefont
  {Xu}}, \bibinfo {author} {\bibfnamefont {J.}~\bibnamefont {Wang}}, \bibinfo
  {author} {\bibfnamefont {H.}~\bibnamefont {Hu}}, \ and\ \bibinfo {author}
  {\bibfnamefont {Y.}~\bibnamefont {Deng}},\ }\bibfield  {title} {\enquote
  {\bibinfo {title} {Critical polynomials in the nonplanar and continuum
  percolation models},}\ }\href {\doibase 10.1103/PhysRevE.103.022127}
  {\bibfield  {journal} {\bibinfo  {journal} {Physical Review E}\ }\textbf
  {\bibinfo {volume} {103}},\ \bibinfo {pages} {022127} (\bibinfo {year}
  {2021})}\BibitemShut {NoStop}%
\bibitem [{\citenamefont {Koza}\ \emph {et~al.}(2014)\citenamefont {Koza},
  \citenamefont {Kondrat},\ and\ \citenamefont
  {Suszczy{\'{n}}ski}}]{Koza_2014}%
  \BibitemOpen
  \bibfield  {author} {\bibinfo {author} {\bibfnamefont {Z.}~\bibnamefont
  {Koza}}, \bibinfo {author} {\bibfnamefont {G.}~\bibnamefont {Kondrat}}, \
  and\ \bibinfo {author} {\bibfnamefont {K.}~\bibnamefont
  {Suszczy{\'{n}}ski}},\ }\bibfield  {title} {\enquote {\bibinfo {title}
  {Percolation of overlapping squares or cubes on a lattice},}\ }\href
  {\doibase 10.1088/1742-5468/2014/11/p11005} {\bibfield  {journal} {\bibinfo
  {journal} {Journal of Statistical Mechanics: Theory and Experiment}\ }\textbf
  {\bibinfo {volume} {2014}},\ \bibinfo {pages} {P11005} (\bibinfo {year}
  {2014})}\BibitemShut {NoStop}%
\bibitem [{\citenamefont {Koza}\ and\ \citenamefont
  {Po{\l}a}(2016)}]{Koza_2016}%
  \BibitemOpen
  \bibfield  {author} {\bibinfo {author} {\bibfnamefont {Z.}~\bibnamefont
  {Koza}}\ and\ \bibinfo {author} {\bibfnamefont {J.}~\bibnamefont {Po{\l}a}},\
  }\bibfield  {title} {\enquote {\bibinfo {title} {From discrete to continuous
  percolation in dimensions 3 to 7},}\ }\href {\doibase
  10.1088/1742-5468/2016/10/103206} {\bibfield  {journal} {\bibinfo  {journal}
  {Journal of Statistical Mechanics: Theory and Experiment}\ }\textbf {\bibinfo
  {volume} {2016}},\ \bibinfo {pages} {103206} (\bibinfo {year}
  {2016})}\BibitemShut {NoStop}%
\bibitem [{\citenamefont {Privman}(1990)}]{Finite-Size_Scaling_Theory_1990}%
  \BibitemOpen
  \bibfield  {author} {\bibinfo {author} {\bibfnamefont {V.}~\bibnamefont
  {Privman}},\ }\enquote {\bibinfo {title} {Finite-size scaling theory},}\ in\
  \href {\doibase 10.1142/9789814503419_0001} {\emph {\bibinfo {booktitle}
  {Finite size scaling and numerical simulation of statistical systems}}},\
  \bibinfo {editor} {edited by\ \bibinfo {editor} {\bibfnamefont
  {V.}~\bibnamefont {Privman}}}\ (\bibinfo  {publisher} {World Scientific},\
  \bibinfo {address} {Singapore},\ \bibinfo {year} {1990})\ pp.\ \bibinfo
  {pages} {1--98}\BibitemShut {NoStop}%
\bibitem [{\citenamefont {Landau}\ and\ \citenamefont
  {Binder}(2005)}]{Guide_to_Monte_Carlo_Simulations_2005}%
  \BibitemOpen
  \bibfield  {author} {\bibinfo {author} {\bibfnamefont {D.~P.}\ \bibnamefont
  {Landau}}\ and\ \bibinfo {author} {\bibfnamefont {K.}~\bibnamefont
  {Binder}},\ }\href {\doibase 10.1017/CBO9780511614460} {\emph {\bibinfo
  {title} {A Guide to Monte Carlo Simulations in Statistical Physics}}},\
  \bibinfo {edition} {2nd}\ ed.\ (\bibinfo  {publisher} {Cambridge University
  Press},\ \bibinfo {year} {2005})\BibitemShut {NoStop}%
\bibitem [{\citenamefont {Newman}\ and\ \citenamefont
  {Ziff}(2001)}]{NewmanZiff2001}%
  \BibitemOpen
  \bibfield  {author} {\bibinfo {author} {\bibfnamefont {M.~E.~J.}\
  \bibnamefont {Newman}}\ and\ \bibinfo {author} {\bibfnamefont {R.~M.}\
  \bibnamefont {Ziff}},\ }\bibfield  {title} {\enquote {\bibinfo {title} {Fast
  {M}onte {C}arlo algorithm for site or bond percolation},}\ }\href {\doibase
  10.1103/PhysRevE.64.016706} {\bibfield  {journal} {\bibinfo  {journal}
  {Physical Review E}\ }\textbf {\bibinfo {volume} {64}},\ \bibinfo {pages}
  {016706} (\bibinfo {year} {2001})}\BibitemShut {NoStop}%
\bibitem [{\citenamefont {Kittel}(2005)}]{Kittel_2005}%
  \BibitemOpen
  \bibfield  {author} {\bibinfo {author} {\bibfnamefont {C.}~\bibnamefont
  {Kittel}},\ }\href@noop {} {\emph {\bibinfo {title} {Introduction to Solid
  State Physics}}},\ \bibinfo {edition} {8th}\ ed.\ (\bibinfo  {publisher}
  {John Wiley \& Sons, Inc},\ \bibinfo {year} {2005})\BibitemShut {NoStop}%
\bibitem [{\citenamefont {Djordjevic}\ \emph {et~al.}(1982)\citenamefont
  {Djordjevic}, \citenamefont {Stanley},\ and\ \citenamefont
  {Margolina}}]{Djordjevic_1982}%
  \BibitemOpen
  \bibfield  {author} {\bibinfo {author} {\bibfnamefont {Z.~V.}\ \bibnamefont
  {Djordjevic}}, \bibinfo {author} {\bibfnamefont {H.~E.}\ \bibnamefont
  {Stanley}}, \ and\ \bibinfo {author} {\bibfnamefont {A.}~\bibnamefont
  {Margolina}},\ }\bibfield  {title} {\enquote {\bibinfo {title} {Site
  percolation threshold for honeycomb and square lattices},}\ }\href {\doibase
  10.1088/0305-4470/15/8/006} {\bibfield  {journal} {\bibinfo  {journal}
  {Journal of Physics A: Mathematical and General}\ }\textbf {\bibinfo {volume}
  {15}},\ \bibinfo {pages} {L405--L412} (\bibinfo {year} {1982})}\BibitemShut
  {NoStop}%
\bibitem [{\citenamefont {Jacobsen}(2014)}]{Jacobsen_2014}%
  \BibitemOpen
  \bibfield  {author} {\bibinfo {author} {\bibfnamefont {J.~L.}\ \bibnamefont
  {Jacobsen}},\ }\bibfield  {title} {\enquote {\bibinfo {title} {High-precision
  percolation thresholds and {P}otts-model critical manifolds from graph
  polynomials},}\ }\href {\doibase 10.1088/1751-8113/47/13/135001} {\bibfield
  {journal} {\bibinfo  {journal} {Journal of Physics A: Mathematical and
  Theoretical}\ }\textbf {\bibinfo {volume} {47}},\ \bibinfo {pages} {135001}
  (\bibinfo {year} {2014})}\BibitemShut {NoStop}%
\bibitem [{\citenamefont {Feng}\ \emph {et~al.}(2008)\citenamefont {Feng},
  \citenamefont {Deng},\ and\ \citenamefont {Bl\"ote}}]{PhysRevE.78.031136}%
  \BibitemOpen
  \bibfield  {author} {\bibinfo {author} {\bibfnamefont {X.}~\bibnamefont
  {Feng}}, \bibinfo {author} {\bibfnamefont {Y.}~\bibnamefont {Deng}}, \ and\
  \bibinfo {author} {\bibfnamefont {H.~W.~J.}\ \bibnamefont {Bl\"ote}},\
  }\bibfield  {title} {\enquote {\bibinfo {title} {Percolation transitions in
  two dimensions},}\ }\href {\doibase 10.1103/PhysRevE.78.031136} {\bibfield
  {journal} {\bibinfo  {journal} {Physical Review E}\ }\textbf {\bibinfo
  {volume} {78}},\ \bibinfo {pages} {031136} (\bibinfo {year}
  {2008})}\BibitemShut {NoStop}%
\end{thebibliography}%

\appendix
\renewcommand\thefigure{X\arabic{figure}} 
\setcounter{figure}{0}

\section{\label{app:functions}Boundaries procedures}

Set of {\tt boundaries()} functions (written in C) to be replaced in the Newman--Ziff program published in Ref. \onlinecite{NewmanZiff2001} to obtain the single realization of $\mathcal S_{\max}(n;L)$ with the honeycomb lattice and the corresponding neighborhoods presented in \Cref{fig:hc-basic-neighbours} and their computerized versions (see \Cref{fig:hc-computerized}).

\subsection{\textsc{hc-1}}
\begin{lstlisting}[language=C]
void boundaries() {
int i,row,col;

for(row=0; row<L; row++) 
for(col=0; col<L; col++) {
   i=row*L+col; 
// 1NN core:
   nn[i][0] = (N+i -1)%N;
   nn[i][1] = (N+i +1)%N;
   if(row%2==0) {
      if(col%2!=0)
         nn[i][2] = (N+i +L)%N;
      else 
         nn[i][2] = (N+i -L)%N; }
   else {
      if(col%2==0)
         nn[i][2] = (N+i +L)%N;
      else 
         nn[i][2] = (N+i -L)%N; }
// 1NN left border:
   if(i%L==0)     
      nn[i][0] = (N+L+i -1)%N;
// 1NN right border:
   if((i+1)%L==0) 
      nn[i][1] = (N-L+i +1)%N;
   }
}
\end{lstlisting}

\subsection{\textsc{hc-2}}
\begin{lstlisting}[language=C]
void boundaries() {
int i,row,col;

for(row=0; row<L; row++) 
for(col=0; col<L; col++) {
   i=row*L+col;
// 2NN core:
   nn[i][0] = (N+i -2  )%N;
   nn[i][1] = (N+i +2  )%N;
   nn[i][2] = (N+i +L-1)%N;
   nn[i][3] = (N+i +L+1)%N;
   nn[i][4] = (N+i -L-1)%N;
   nn[i][5] = (N+i -L+1)%N; 
// 2NN left border:
   if(i%L==0) {
      nn[i][0] = (N+L+i -2  )%N;
      nn[i][2] = (N+L+i +L-1)%N;
      nn[i][4] = (N+L+i -L-1)%N; }
   if(i%L==1) {
      nn[i][0] = (N+L+i -2  )%N; }
// 2NN right border:
   if((i+1)%L==0) {
      nn[i][1] = (N-L+i +2  )%N;
      nn[i][3] = (N-L+i +L+1)%N;
      nn[i][5] = (N-L+i -L+1)%N; }
   if((i+2)%L==0) {
      nn[i][1] = (N-L+i +2  )%N; }
   }
}
\end{lstlisting}

\subsection{\textsc{hc-3}}
\begin{lstlisting}[language=C]
void boundaries() {
int i,row,col;

for(row=0; row<L; row++)
for(col=0; col<L; col++) {
   i=row*L+col;
// 3NN core:
   if(row%2==0) {
      if(col%2==0) {
         nn[i][0] = (N+i +L)%N;
         nn[i][1] = (N+i -L-2)%N;
         nn[i][2] = (N+i -L+2)%N; }
      else {
         nn[i][0] = (N+i -L)%N;
         nn[i][1] = (N+i +L-2)%N;
         nn[i][2] = (N+i +L+2)%N; }
      } 
   else {
      if(col%2==0) {
         nn[i][0] = (N+i -L)%N;
         nn[i][1] = (N+i +L-2)%N;
         nn[i][2] = (N+i +L+2)%N; }
      else {
         nn[i][0] = (N+i +L)%N;
         nn[i][1] = (N+i -L-2)%N;
         nn[i][2] = (N+i -L+2)%N; }
   }
// 3NN left border:
   if(i%L==0 || i%L==1) {
      if(row%2==0) {
         if(col%2==0)
            nn[i][1] = (N+L+i -L-2)%N;
         else
            nn[i][1] = (N+L+i +L-2)%N; }
      else {
         if(col%2==0) 
            nn[i][1] = (N+L+i +L-2)%N;
         else 
            nn[i][1] = (N+L+i -L-2)%N; }
   }
// 3NN right border:
   if((i+1)%L==0 || (i+2)%L==0) {
     if(row%2==0) {
       if(col%2==0)
         nn[i][2] = (N-L+i -L+2)%N;
       else
         nn[i][2] = (N-L+i +L+2)%N; }
      else {
        if(col%2==0)
          nn[i][2] = (N-L+i +L+2)%N;
        else
          nn[i][2] = (N-L+i -L+2)%N; }
   }
   }
}
\end{lstlisting}

\subsection{\textsc{hc-4}}
\begin{lstlisting}[language=C]
void boundaries() {
int i,row,col;

for(row=0; row<L; row++)
for(col=0; col<L; col++) {
   i=row*L+col;
// 4NN core:
   nn[i][0] = (N+i -3)%N;
   nn[i][3] = (N+i +3)%N;
   if(row%2==0) {
      if(col%2==0) {
         nn[i][1] = (N+i +L-2)%N;
         nn[i][2] = (N+i -2*L-1)%N;
         nn[i][4] = (N+i +L+2)%N;
         nn[i][5] = (N+i -2*L+1)%N; }
      else {
         nn[i][1] = (N+i +2*L-1)%N;
         nn[i][2] = (N+i -L-2)%N;
         nn[i][4] = (N+i +2*L+1)%N;
         nn[i][5] = (N+i -L+2)%N; }
   }
   else {
      if(col%2==0) {
         nn[i][1] = (N+i +2*L-1)%N;
         nn[i][2] = (N+i -L-2)%N;
         nn[i][4] = (N+i +2*L+1)%N;
         nn[i][5] = (N+i -L+2)%N; }
      else {
         nn[i][1] = (N+i +L-2)%N;
         nn[i][2] = (N+i -2*L-1)%N;
         nn[i][4] = (N+i +L+2)%N;
         nn[i][5] = (N+i -2*L+1)%N; }
   } 
// 4NN left border:
   if(i%L==0 || i%L==1 || i%L==2)
      nn[i][0] = (N+L+i -3)%N;
   if(i%L==0 || i%L==1)
      if(row%2==0) {
      if(col%2==0)
         nn[i][1] = (N+L+i +L-2)%N;
      else
         nn[i][2] = (N+L+i -L-2)%N;
      } 
   else {
      if(col%2==0)
         nn[i][2] = (N+L+i -L-2)%N;
      else
         nn[i][1] = (N+L+i +L-2)%N;
   }
   if(i%L==0)
      if(row%2==0) {
      if(col%2==0) 
         nn[i][2] = (N+L+i -2*L-1)%N;
      else
         nn[i][1] = (N+L+i +2*L-1)%N;
      }
   else {
      if(col%2==0) 
         nn[i][1] = (N+L+i +2*L-1)%N;
      else
         nn[i][2] = (N+L+i -2*L-1)%N;
   } 
// 4NN right border:
   if((i+1)%L==0 || (i+2)%L==0 || (i+3)%L==0)
      nn[i][3] = (N-L+i +3)%N;
   if((i+1)%L==0 || (i+2)%L==0)
      if(row%2==0) {
      if(col%2==0)
         nn[i][4] = (N-L+i +L+2)%N;
      else
         nn[i][5] = (N-L+i -L+2)%N;
      }
   else {
     if(col%2==0)
       nn[i][5] = (N-L+i -L+2)%N;
     else
       nn[i][4] = (N-L+i +L+2)%N;
   } 
   if((i+1)%L==0)
     if(row%2==0) {
     if(col%2==0)
        nn[i][5] = (N-L+i -2*L+1)%N;
     else
        nn[i][4] = (N-L+i +2*L+1)%N;
     }
   else {
      if(col%2==0)
         nn[i][4] = (N-L+i +2*L+1)%N;
      else
         nn[i][5] = (N-L+i -2*L+1)%N;
   }
}
}
\end{lstlisting}

\subsection{\textsc{hc-5}}
\begin{lstlisting}[language=C]
void boundaries() {
int i,row,col;

for(row=0; row<L; row++) 
for(col=0; col<L; col++) {
   i=row*L+col;
// 5NN core:
   nn[i][0] = (N+i -2*L)%N;
   nn[i][1] = (N+i +2*L)%N;
   nn[i][2] = (N+i +L-3)%N;
   nn[i][3] = (N+i +L+3)%N;
   nn[i][4] = (N+i -L-3)%N;
   nn[i][5] = (N+i -L+3)%N; 
// 5NN left border:
   if(i%L==0 || i%L==1 || i%L==2) {
      nn[i][2] = (N+L+i +L-3)%N;
      nn[i][4] = (N+L+i -L-3)%N; }
// 5NN right border:
   if((i+1)%L==0 || (i+2)%L==0 || (i+3)%L==0) {
      nn[i][3] = (N-L+i +L+3)%N;
      nn[i][5] = (N-L+i -L+3)%N; }
}
}
\end{lstlisting}

\section{\label{app:hc-neighbours}Complex neighborhoods shapes}

\begin{figure*}[htbp]
\input{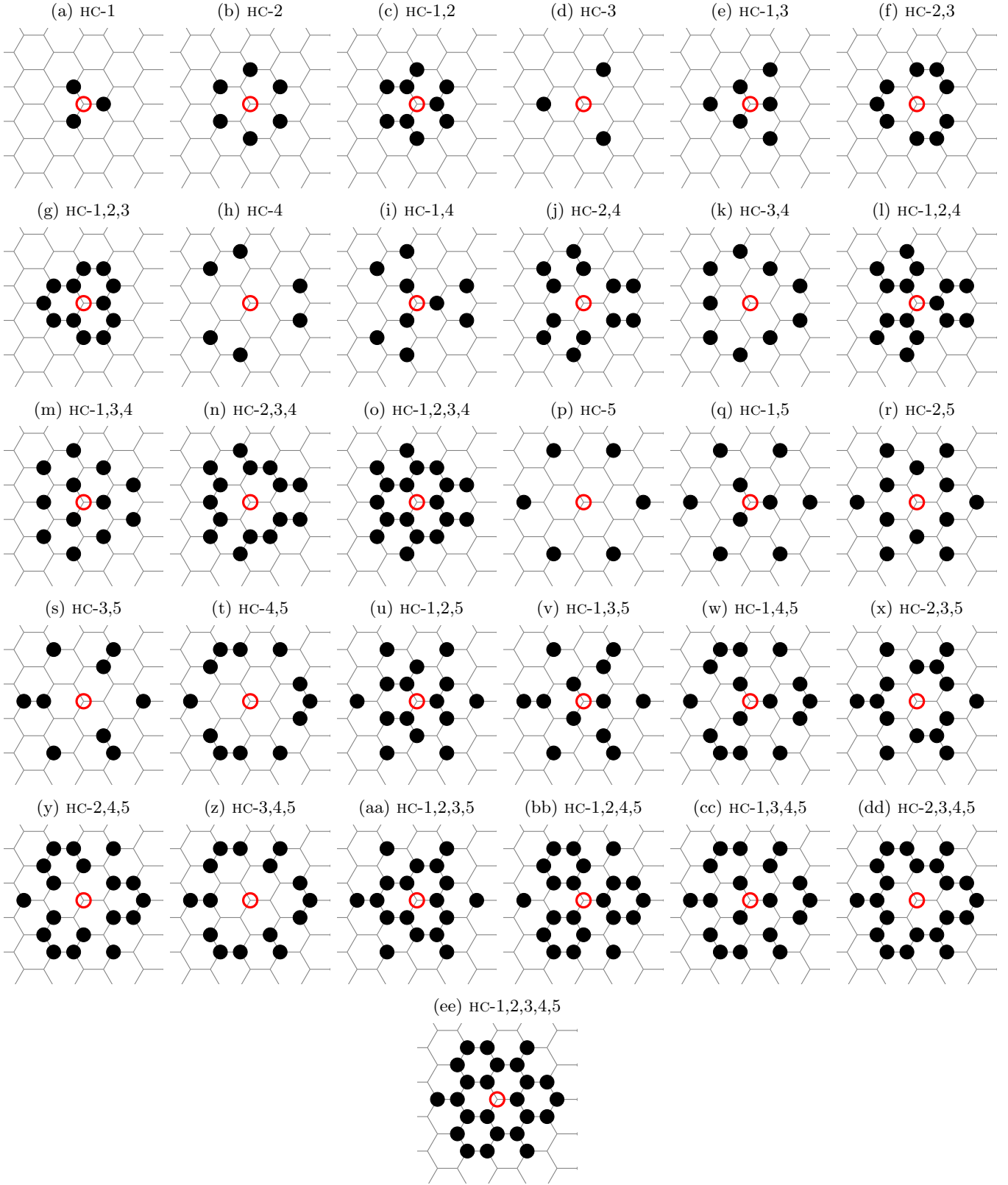}
\caption{\label{fig:hc-neighbours}Complex neighborhoods in a honeycomb lattice}
\end{figure*}

In \Cref{fig:hc-neighbours} 31 complex neighborhoods in a honeycomb lattice are presented.
The central site is marked by a red open circle, whereas the sites in its neighborhoods are marked by solid black circles.
The names of the neighborhoods are presented in subfigure headlines. 

\section{\label{app:PmaxLbetanuvsp}Finite-size scaling}

In \Cref{fig:PmaxLbetanuvsp-various} we show $\mathcal{P}_{\max}\cdot L^{5/48}$ versus $p$ for neighborhoods containing sites up to the fifth coordination zone.
The names of the neighborhoods are presented in subfigure headlines.

\begin{figure*}[htbp]
\begin{subfigure}[b]{0.32\textwidth}
\caption{\label{fig:PmaxLbetanuvsp-1}\textsc{hc}-1}
\includegraphics[width=0.99\textwidth]{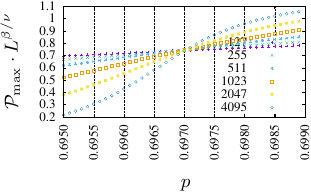}
\end{subfigure}
\begin{subfigure}[b]{0.32\textwidth}
\caption{\label{fig:PmaxLbetanuvsp-2}\textsc{hc}-2}
\includegraphics[width=0.99\textwidth]{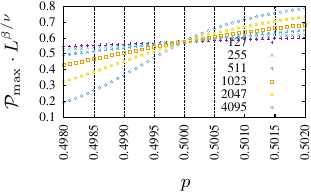}
\end{subfigure}
\begin{subfigure}[b]{0.32\textwidth}
\caption{\label{fig:PmaxLbetanuvsp-12}\textsc{hc}-1,2}
\includegraphics[width=0.99\textwidth]{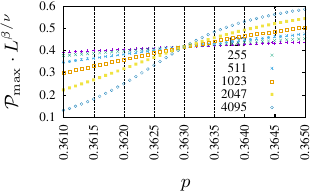}
\end{subfigure}
\begin{subfigure}[b]{0.32\textwidth}
\caption{\label{fig:PmaxLbetanuvsp-3}\textsc{hc}-3}
\includegraphics[width=0.99\textwidth]{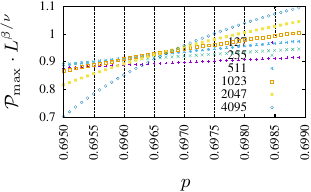}
\end{subfigure}
\begin{subfigure}[b]{0.32\textwidth}
\caption{\label{fig:PmaxLbetanuvsp-13}\textsc{hc}-1,3}
\includegraphics[width=0.99\textwidth]{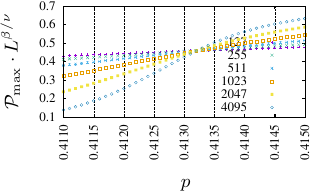}
\end{subfigure}
\begin{subfigure}[b]{0.32\textwidth}
\caption{\label{fig:PmaxLbetanuvsp-23}\textsc{hc}-2,3}
\includegraphics[width=0.99\textwidth]{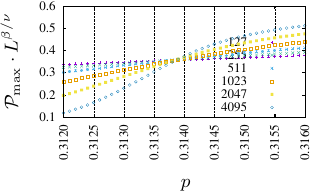}
\end{subfigure}
\begin{subfigure}[b]{0.32\textwidth}
\caption{\label{fig:PmaxLbetanuvsp-123}\textsc{hc}-1,2,3}
\includegraphics[width=0.99\textwidth]{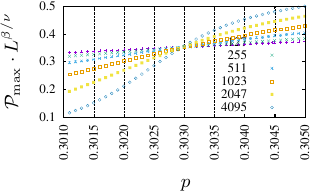}
\end{subfigure}
\begin{subfigure}[b]{0.32\textwidth}
\caption{\label{fig:PmaxLbetanuvsp-4}\textsc{hc}-4}
\includegraphics[width=0.99\textwidth]{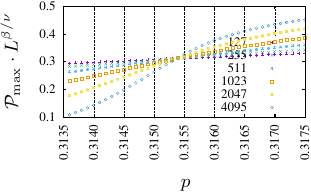}
\end{subfigure}
\begin{subfigure}[b]{0.32\textwidth}
\caption{\label{fig:PmaxLbetanuvsp-14}\textsc{hc}-1,4}
\includegraphics[width=0.99\textwidth]{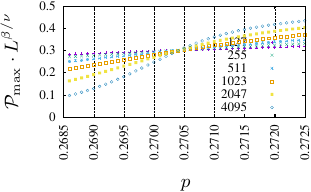}
\end{subfigure}
\begin{subfigure}[b]{0.32\textwidth}
\caption{\label{fig:PmaxLbetanuvsp-24}\textsc{hc}-2,4}
\includegraphics[width=0.99\textwidth]{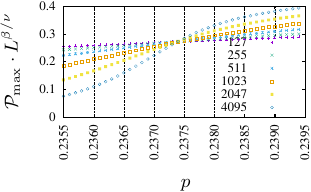}
\end{subfigure}
\begin{subfigure}[b]{0.32\textwidth}
\caption{\label{fig:PmaxLbetanuvsp-34}\textsc{hc}-3,4}
\includegraphics[width=0.99\textwidth]{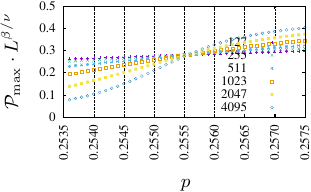}
\end{subfigure}
\begin{subfigure}[b]{0.32\textwidth}
\caption{\label{fig:PmaxLbetanuvsp-124}\textsc{hc}-1,2,4}
\includegraphics[width=0.99\textwidth]{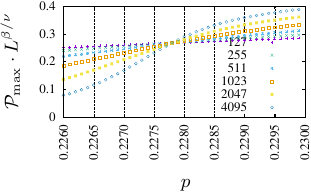}
\end{subfigure}
\begin{subfigure}[b]{0.32\textwidth}
\caption{\label{fig:PmaxLbetanuvsp-134}\textsc{hc}-1,3,4}
\includegraphics[width=0.99\textwidth]{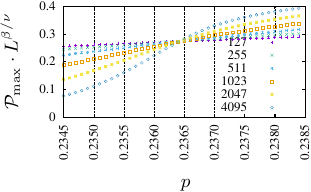}
\end{subfigure}
\begin{subfigure}[b]{0.32\textwidth}
\caption{\label{fig:PmaxLbetanuvsp-234}\textsc{hc}-2,3,4}
\includegraphics[width=0.99\textwidth]{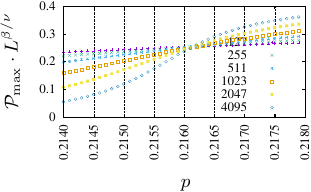}
\end{subfigure}
\begin{subfigure}[b]{0.32\textwidth}
\caption{\label{fig:PmaxLbetanuvsp-1234}\textsc{hc}-1,2,3,4}
\includegraphics[width=0.99\textwidth]{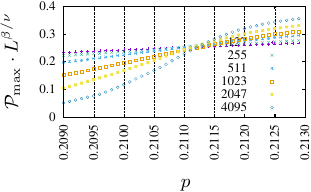}
\end{subfigure}
\begin{subfigure}[b]{0.32\textwidth}
\caption{\label{fig:PmaxLbetanuvsp-5}\textsc{hc}-5}
\includegraphics[width=0.99\textwidth]{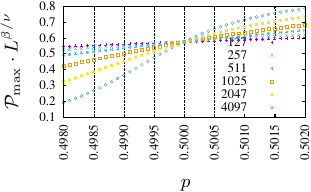}
\end{subfigure}
\begin{subfigure}[b]{0.32\textwidth}
\caption{\label{fig:PmaxLbetanuvsp-15}\textsc{hc}-1,5}
\includegraphics[width=0.99\textwidth]{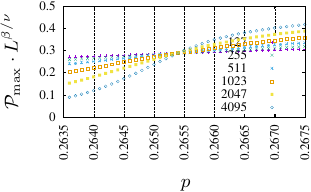}
\end{subfigure}
\begin{subfigure}[b]{0.32\textwidth}
\caption{\label{fig:PmaxLbetanuvsp-25}\textsc{hc}-2,5}
\includegraphics[width=0.99\textwidth]{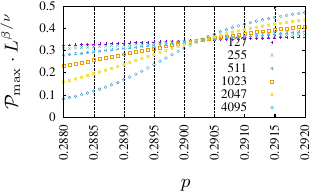}
\end{subfigure}
\end{figure*}


\begin{figure*}\ContinuedFloat
\begin{subfigure}[b]{0.32\textwidth}
\caption{\label{fig:PmaxLbetanuvsp-35}\textsc{hc}-3,5}
\includegraphics[width=0.99\textwidth]{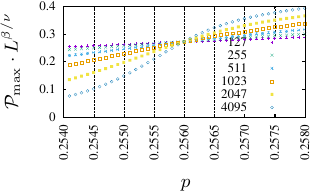}
\end{subfigure}
\begin{subfigure}[b]{0.32\textwidth}
\caption{\label{fig:PmaxLbetanuvsp-45}\textsc{hc}-4,5}
\includegraphics[width=0.99\textwidth]{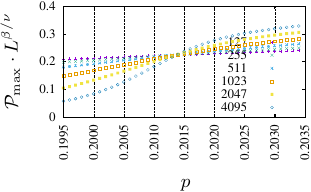}
\end{subfigure}
\begin{subfigure}[b]{0.32\textwidth}
\caption{\label{fig:PmaxLbetanuvsp-125}\textsc{hc}-1,2,5}
\includegraphics[width=0.99\textwidth]{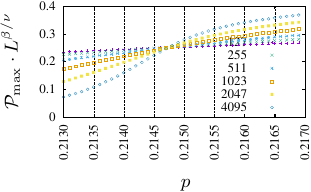}
\end{subfigure}
\begin{subfigure}[b]{0.32\textwidth}
\caption{\label{fig:PmaxLbetanuvsp-135}\textsc{hc}-1,3,5}
\includegraphics[width=0.99\textwidth]{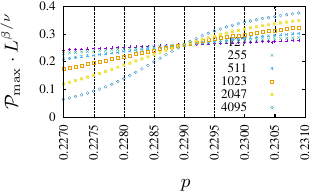}
\end{subfigure}
\begin{subfigure}[b]{0.32\textwidth}
\caption{\label{fig:PmaxLbetanuvsp-145}\textsc{hc}-1,4,5}
\includegraphics[width=0.99\textwidth]{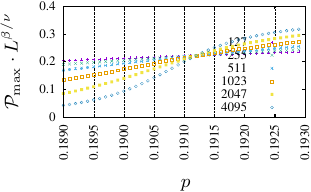}
\end{subfigure}
\begin{subfigure}[b]{0.32\textwidth}
\caption{\label{fig:PmaxLbetanuvsp-235}\textsc{hc}-2,3,5}
\includegraphics[width=0.99\textwidth]{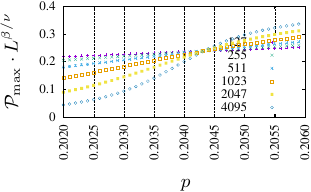}
\end{subfigure}
\begin{subfigure}[b]{0.32\textwidth}
\caption{\label{fig:PmaxLbetanuvsp-245}\textsc{hc}-2,4,5}
\includegraphics[width=0.99\textwidth]{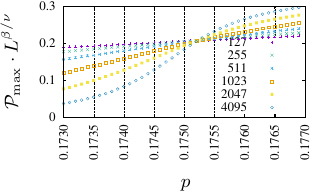}
\end{subfigure}
\begin{subfigure}[b]{0.32\textwidth}
\caption{\label{fig:PmaxLbetanuvsp-345}\textsc{hc}-3,4,5}
\includegraphics[width=0.99\textwidth]{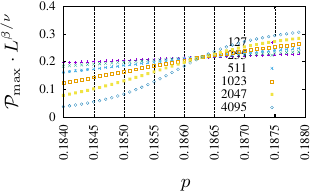}
\end{subfigure}
\begin{subfigure}[b]{0.32\textwidth}
\caption{\label{fig:PmaxLbetanuvsp-1235}\textsc{hc}-1,2,3,5}
\includegraphics[width=0.99\textwidth]{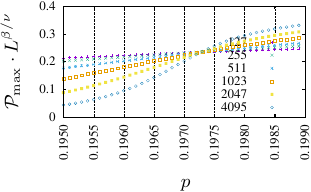}
\end{subfigure}
\begin{subfigure}[b]{0.32\textwidth}
\caption{\label{fig:PmaxLbetanuvsp-1245}\textsc{hc}-1,2,4,5}
\includegraphics[width=0.99\textwidth]{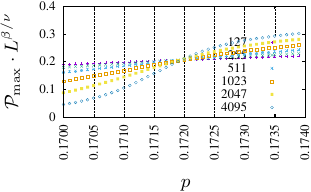}
\end{subfigure}
\begin{subfigure}[b]{0.32\textwidth}
\caption{\label{fig:PmaxLbetanuvsp-1345}\textsc{hc}-1,3,4,5}
\includegraphics[width=0.99\textwidth]{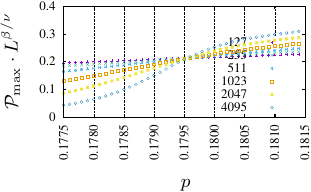}
\end{subfigure}
\begin{subfigure}[b]{0.32\textwidth}
\caption{\label{fig:PmaxLbetanuvsp-2345}\textsc{hc}-2,3,4,5}
\includegraphics[width=0.99\textwidth]{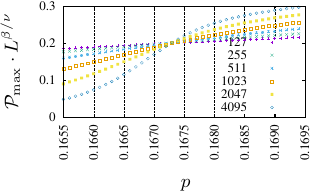}
\end{subfigure}
\begin{subfigure}[b]{0.32\textwidth}
\caption{\label{fig:PmaxLbetanuvsp-12345}\textsc{hc}-1,2,3,4,5}
\includegraphics[width=0.99\textwidth]{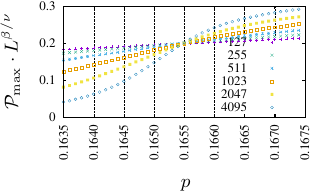}
\end{subfigure}
\caption{\label{fig:PmaxLbetanuvsp-various}(Color online). $\mathcal{P}_{\max}\cdot L^{5/48}$ vs. $p$ for the honeycomb lattice and for the neighborhoods containing sites up to the fifth coordination zone}
\end{figure*}

\end{document}